\documentclass[times,twocolumn]{aastex63}
\usepackage[utf8]{inputenc}
\usepackage[normalem]{ulem}
\usepackage{amsmath}
\usepackage{array,multirow,graphicx}
\usepackage{verbatim}
\usepackage{upgreek}

\newcommand{\HII}{H\,{\sc ii}\ }
\newcommand{\DM}{{\rm DM}}

\newcommand{\be}{\begin{eqnarray}}
\newcommand{\ee}{\end{eqnarray}}

\begin{document}

\title{Fast Radio Bursts Co-detected by CHIME/FRB and
DSA-110 from 400 to 1530 MHz} 

\author[0000-0002-4941-5333]{Stella Koch Ocker}
\altaffiliation{These authors contributed equally to this work}
\affiliation{The Observatories of the Carnegie Institution for Science, Pasadena, CA 91101}
\affiliation{Cahill Center for Astronomy and Astrophysics, California Institute of Technology, Pasadena, CA 91125}

\author[0000-0003-0510-0740]{Kenzie Nimmo}
\altaffiliation{These authors contributed equally to this work}
\affiliation{Center for Astrophysics $|$ Harvard \& Smithsonian, 60 Garden St, Cambridge, MA 02138, USA}
\affiliation{NASA Einstein Fellow}

\author[0000-0001-9855-5781]{Jakob Faber}
\affiliation{Cahill Center for Astronomy and Astrophysics, California Institute of Technology, Pasadena, CA 91125}

\author[0000-0002-6823-2073]{Kaitlyn Shin}
\affiliation{Cahill Center for Astronomy and Astrophysics, California Institute of Technology, Pasadena, CA 91125}

\author[0000-0002-7252-5485]{Vikram Ravi}
\affiliation{Cahill Center for Astronomy and Astrophysics, California Institute of Technology, Pasadena, CA 91125}
\affiliation{Owens Valley Radio Observatory, California Institute of Technology, 100 Leighton Lane, Big Pine, CA 93513, USA}

\author[0009-0005-3565-4164]{Frank M. Verdi}
\affiliation{Cahill Center for Astronomy and Astrophysics, California Institute of Technology, Pasadena, CA 91125}

\author[0009-0009-0938-1595]{Naman Jain}
\affiliation{Department of Physics, McGill University, 3600 rue University, Montr\'eal, QC H3A 2T8, Canada}
\affiliation{Trottier Space Institute, McGill University, 3550 rue University, Montr\'eal, QC H3A 2A7, Canada}

\author[0000-0002-4795-697X]{Ziggy Pleunis}
\affiliation{Anton Pannekoek Institute for Astronomy, University of Amsterdam, Science Park 904, 1098 XH Amsterdam, The Netherlands}
\affiliation{ASTRON, Netherlands Institute for Radio Astronomy, Oude Hoogeveensedijk 4, 7991 PD Dwingeloo, The Netherlands}

\author[0000-0002-8897-1973]{Ayush Pandhi}
\affiliation{Department of Physics, McGill University, 3600 rue University, Montr\'eal, QC H3A 2T8, Canada}
\affiliation{Trottier Space Institute, McGill University, 3550 rue University, Montr\'eal, QC H3A 2A7, Canada}

\author[0000-0002-6573-7316]{Myles Sherman}
\affiliation{Cahill Center for Astronomy and Astrophysics, California Institute of Technology, Pasadena, CA 91125}

\author[0000-0003-3236-8769]{Dylan Jow}
\affiliation{Kavli Institute for Particle Astrophysics \& Cosmology, P.O. Box 2450, Stanford University, Stanford, CA 94305, USA}

\author[0009-0008-2000-6959]{Sachin Pradeep E.T.}
\affiliation{Department of Physics, McGill University, 3600 rue University, Montr\'eal, QC H3A 2T8, Canada}
\affiliation{Trottier Space Institute, McGill University, 3550 rue University, Montr\'eal, QC H3A 2A7, Canada}

\author[0009-0005-5370-7653]{Daniel	Amouyal}
\affiliation{Department of Physics, McGill University, 3600 rue University, Montr\'eal, QC H3A 2T8, Canada}
\affiliation{Trottier Space Institute, McGill University, 3550 rue University, Montr\'eal, QC H3A 2A7, Canada}

\author[0000-0002-7587-6352]{Liam	Connor}
\affiliation{Center for Astrophysics $|$ Harvard \& Smithsonian, 60 Garden St, Cambridge, MA 02138, USA}

\author[0000-0002-8376-1563]{Alice P.	Curtin}
\affiliation{Department of Physics, McGill University, 3600 rue University, Montr\'eal, QC H3A 2T8, Canada}
\affiliation{Trottier Space Institute, McGill University, 3550 rue University, Montr\'eal, QC H3A 2A7, Canada}

\author[0009-0006-1258-4228]{Maxwell Fine}
\affiliation{Department of Physics, McGill University, 3600 rue University, Montr\'eal, QC H3A 2T8, Canada}
\affiliation{Trottier Space Institute, McGill University, 3550 rue University, Montr\'eal, QC H3A 2A7, Canada}

\author[0000-0002-3382-9558]{B. M.	Gaensler}
\affiliation{Department of Astronomy and Astrophysics, University of California, Santa Cruz, 1156 High Street, Santa Cruz, CA 95060, USA}
\affiliation{Dunlap Institute for Astronomy and Astrophysics, 50 St. George Street, University of Toronto, ON M5S 3H4, Canada}
\affiliation{David A.\ Dunlap Department of Astronomy and Astrophysics, 50 St. George Street, University of Toronto, ON M5S 3H4, Canada}

\author[0000-0003-2317-1446]{Jason W.~T. Hessels}
\affiliation{Department of Physics, McGill University, 3600 rue University, Montr\'eal, QC H3A 2T8, Canada}
\affiliation{Trottier Space Institute, McGill University, 3550 rue University, Montr\'eal, QC H3A 2A7, Canada}
\affiliation{Anton Pannekoek Institute for Astronomy, University of Amsterdam, Science Park 904, 1098 XH Amsterdam, The Netherlands}
\affiliation{ASTRON, Netherlands Institute for Radio Astronomy, Oude Hoogeveensedijk 4, 7991 PD Dwingeloo, The Netherlands}

\author[0000-0001-9345-0307]{Victoria Kaspi}
\affiliation{Department of Physics, McGill University, 3600 rue University, Montr\'eal, QC H3A 2T8, Canada}
\affiliation{Trottier Space Institute, McGill University, 3550 rue University, Montr\'eal, QC H3A 2A7, Canada}

\author[0009-0004-4176-0062]{Afrokk	Khan}
\affiliation{Department of Physics, McGill University, 3600 rue University, Montr\'eal, QC H3A 2T8, Canada}
\affiliation{Trottier Space Institute, McGill University, 3550 rue University, Montr\'eal, QC H3A 2A7, Canada}

\author[0000-0002-4119-9963]{Casey	Law}
\affiliation{Cahill Center for Astronomy and Astrophysics, California Institute of Technology, Pasadena, CA 91125}
\affiliation{Owens Valley Radio Observatory, California Institute of Technology, 100 Leighton Lane, Big Pine, CA 93513, USA}

\author[0000-0002-7164-9507]{Robert	Main}
\affiliation{Dominion Radio Astrophysical Observatory, Herzberg Astronomy \& Astrophysics Research Centre, National Research Council Canada, 717 White Lake Rd., Kaleden, BC, V0H 1K0 Canada}

\author[0000-0002-0786-7307]{Kyle	McGregor}
\affiliation{Department of Physics, McGill University, 3600 rue University, Montr\'eal, QC H3A 2T8, Canada}
\affiliation{Trottier Space Institute, McGill University, 3550 rue University, Montr\'eal, QC H3A 2A7, Canada}

\author[0000-0002-8912-0732]{Aaron~B.~Pearlman}
\altaffiliation{NASA Hubble Fellow.}
\affiliation{MIT Kavli Institute for Astrophysics and Space Research, Massachusetts Institute of Technology, 77 Massachusetts Avenue, Cambridge, MA 02139, USA}
\affiliation{Department of Physics, McGill University, 3600 rue University, Montr\'eal, QC H3A 2T8, Canada}
\affiliation{Trottier Space Institute, McGill University, 3550 rue University, Montr\'eal, QC H3A 2A7, Canada}

\author[0000-0002-4623-5329]{Mawson	Sammons}
\affiliation{Laboratoire d'Astrophysique de Marseille, Aix-Marseille Univ., CNRS, CNES, Marseille, France}

\author[0000-0002-7374-7119]{Paul	Scholz}
\affiliation{Department of Physics and Astronomy, York University, 4700 Keele Street, Toronto, ON MJ3 1P3, Canada}

\author[0000-0002-4477-3625]{Kritti 	Sharma}
\affiliation{Cahill Center for Astronomy and Astrophysics, California Institute of Technology, Pasadena, CA 91125}

\author[0000-0002-2088-3125]{Kendrick	Smith}
\affiliation{Perimeter Institute for Theoretical Physics, 31 Caroline Street N, Waterloo, ON N25 2YL, Canada}

\author[0000-0001-9784-8670]{Ingrid	Stairs}
\affiliation{Department of Physics and Astronomy, University of British Columbia, 6224 Agricultural Road, Vancouver, BC V6T 1Z1 Canada}


\begin{abstract}
    We present a sample of $15$ fast radio bursts (FRBs) that were co-detected by the Deep Synoptic Array (DSA-110) and the Canadian Hydrogen Intensity Mapping Experiment (CHIME) between 2022 and 2024, of which $12$ have channelized voltage data from both instruments and 9 have secure redshifts. All but one are as-yet-non-repeating.
    Most bursts span the entire bandwidth of each instrument (400--800\,MHz for CHIME, 1280--1530\,MHz for DSA-110), significantly expanding the number of FRBs known to emit across $>1$\,GHz of spectral bandwidth and constraining the spectral evolution of energy and propagation effects. On average, the FRBs exhibit higher spectral energies (up to a factor of $\sim10$) in the CHIME band but are lower in peak flux density due to substantial pulse broadening. The sample mean (median) spectral index $-0.4\pm0.4$  ($-0.3\pm0.4$) is shallower than previously reported spectral indices, likely a selection effect. A third of the sample shows evidence for a spectral turnover, comparable to that seen in GHz-peaked pulsar and magnetar spectra. 
    Joint scintillation and pulse broadening constraints indicate that nearly all bursts encounter multiple scattering screens, with pulse broadening originating within their host galaxies. The co-detected bursts are among the most scattered in the CHIME/FRB sample, while being the narrowest in the DSA-110 sample, indicating that CHIME/FRB and DSA-110 select different subsets of the FRB population; the interplay between intrinsic burst width, scattering, and the FRB search time resolution creates a narrow window of sensitivity to co-detected FRBs, providing a blueprint for future ultra-wideband FRB surveys.
\end{abstract}


\section{Introduction}

Fast radio bursts \citep[FRBs;][]{lorimer2007,thornton2013} are compact, energetic radio sources that probe the extreme ends of stellar evolution and act as backlights through ionized gas on Galactic, circumgalactic, and intergalactic scales. Their radio emission undergoes propagation effects that are sensitive to key properties of plasma intervening their lines-of-sight (LOSs), including the electron column density, or dispersion measure (DM), and the electron column density weighted by the LOS magnetic field, or Faraday rotation measure (RM).
Radio waves also undergo multipath propagation through electron density fluctuations, which yield a range of observable scattering effects including scintillation and pulse broadening. Accurately disentangling extrinsic propagation effects from FRBs' intrinsic properties (i.e., those that are imprinted in the emission process) is vital for understanding the origins of these extreme radio sources.

Many of these propagation effects and emission signatures are degenerate when observed within a  narrow frequency bandwidth. Pulse broadening might mimic an intrinsically wide burst or multiple blended components, and scintillation can be confused with intrinsic spectral structure, instrumental correlations, or self-noise, i.e. the random intensity fluctuations inherent to measuring a noise-like signal at finite time-frequency resolution. A single frequency-band snapshot of a burst's spectrum also limits constraints on its total energy. The literature commonly assumes a fixed $1$\,GHz spectral extent when deriving FRB energies, despite FRBs typically not being detected across that range \citep[e.g.][]{james2022, shin2022,Shah_2026_arXiv}. Narrowband observations also limit the ability to identify induced spectral structure, such as free-free absorption from the local environment (e.g. seen in some pulsars and magnetars; \citealt{kramer2003, Kijak_2011_A&A, Maan_2022_ApJ}), plasma lensing \citep{uttarkar2026}, or Faraday conversion \citep[][]{lower2024}. 

Ultra-wideband observations offer a direct way to break these degeneracies. The strong inverse frequency dependence of the pulse broadening delay ($\tau\propto\nu^{-\alpha}$) and scintillation bandwidth ($\Delta\nu\propto\nu^{\alpha}$, with $\alpha\approx4$ typically) means that wide frequency coverage allows both effects to be measured simultaneously, which enables localization of scattering structures along the LOS \citep[e.g.][]{2015Natur.528..523M,ocker2021halo,sammons2023}. Scattering screens along FRB sightlines have also been localized using comparatively narrowband data, through the temporal variation of scintillation velocities and arcs \citep{Main_2023_MNRAS, Wu_2024_ApJL}, though this technique requires repeated bursts over long timescales and is thus not possible for apparently non-repeating FRBs.
Ultra-wideband receivers \citep[e.g.][]{hobbs_parkes_uwl,lynch2023} additionally make it possible to explore just how broadband FRB emission can be, and to use a burst's measured spectral shape, rather than an extrapolation beyond the telescope's own bandwidth, to constrain its isotropic-equivalent energy. 

To date, FRB studies over very wide frequency bands have been limited to targeted use of telescopes with small fields-of-view, whether through observations with a single ultra-wideband receiver \citep{kumar2023,uttarkar2026,limaye2026} or through coordinated simultaneous observing with multiple receivers \citep{Majid_2020_ApJL,wang2025_FRB20240114A,bhattacharyya2026}, sometimes distributed across the globe \citep{Law_2017_ApJ,Pearlman_2020_ApJL, pastor2021, Kirsten_2024_NatAs, Ould-Boukattine_2026_arXiv}. All of these observations have necessarily focused on repeating FRB sources, whose spectra also happen to have narrower frequency bandwidths, on average, than those of apparently non-repeating sources \citep{pleunis2021}, leaving the wideband behaviour of the non-repeating majority of the population ($\sim97$\%; \citealt{Cook_2026_arXiv}) almost entirely unexplored.

Extant FRB surveys at disparate frequency bands with overlapping sky coverage offer an immediate opportunity to extend wideband studies to non-repeating sources, ahead of the ultra-wideband surveys soon to come online (e.g., the Canadian Hydrogen Observatory and Radio-transient Detector, CHORD, and the Deep Synoptic Array, DSA; \citealt{chord_overview,connor2026}). In this study, we present a sample of 15 FRBs detected by both the Canadian Hydrogen Intensity Mapping Experiment (CHIME; located at the Dominion Radio Astrophysical Observatory in British Columbia) and the DSA-110 (at Owens Valley Radio Observatory in California) between 2022 February and 2024 February, firmly establishing that apparently non-repeating FRBs can emit across more than 1\,GHz of spectral bandwidth. During this time period, DSA-110 shared $<5\%$ of CHIME's sky coverage, at distinct frequency bands: CHIME at 400--800\,MHz and DSA-110 at 1280--1530\,MHz. Of the 15 bursts that were detected by both telescopes, 12 have data with sufficient signal-to-noise (S/N) and spectro-temporal resolution for a detailed assessment of their spectral morphologies and propagation effects. This paper presents the co-detected sample of FRBs, analyzes their spectral morphologies and energies, and characterizes the scattering along their LOSs. In a companion paper (Pandhi et al., in prep, hereafter referred to as ``Paper II"), we characterize the spectral polarization of these FRBs and its relation to scattering, and we interpret the results in the context of FRB emission mechanisms and local environments. 

This co-detected sample demonstrates the rich scientific return of wideband FRB observations, in advance of the orders of magnitude more wideband FRBs to be detected by CHORD and DSA. Section\,\ref{sec:sample} introduces the CHIME and DSA-110 telescopes, specifies our conditions for a burst to be co-detected by both telescopes, and gives an overview of the co-detected sample. In Section\,\ref{sec:spectra} we characterize the spectral morphology of each burst, highlighting evidence for a spectral turnover in some bursts. Further in this section we measure isotropic-equivalent burst energies based on evidence that the bursts span the full 1.1 GHz spectral extent of both instruments combined. Scattering measurements, including pulse broadening and scintillation, are presented in Section~\ref{sec:scattering-analysis}: 9 bursts show evidence for two distinct scattering screens along their LOS. 
These measurements are used to deconstruct the distribution of scattering material along the LOSs in Section\,\ref{sec:screens}. In Section\,\ref{sec:discussion} we discuss the implications of these results, before concluding in Section\,\ref{sec:conclusion}. Throughout the paper we adopt \cite{planck18} cosmological parameters.

\textit{Nomenclature:} The FRBs presented here were independently reported to the Transient Name Server (TNS) by the DSA-110 and CHIME/FRB as unique sources. Both TNS names are listed for each source in Table~\ref{tab:burst-props}; throughout the paper we refer to each FRB by the date on which it was detected and both letters recorded in TNS for each FRB (CHIME/FRB letter followed by  DSA-110).\footnote{The two TNS entries for each burst will be linked to indicate a common source.}

\section{The Co-detection Sample}\label{sec:sample}

CHIME and DSA-110 are highly distinct in their instrumental design and survey strategy; nonetheless their substantial overlap in sky coverage and high detection rates make them suited for independent detection of identical sources. Overviews of each telescope and their FRB surveys can be found in \cite{chimecat1,chimecat2,ravi2023,law2024}; here we emphasize their characteristics most relevant to this work.

\begin{figure}
    \centering
    \includegraphics[width=0.45\textwidth]{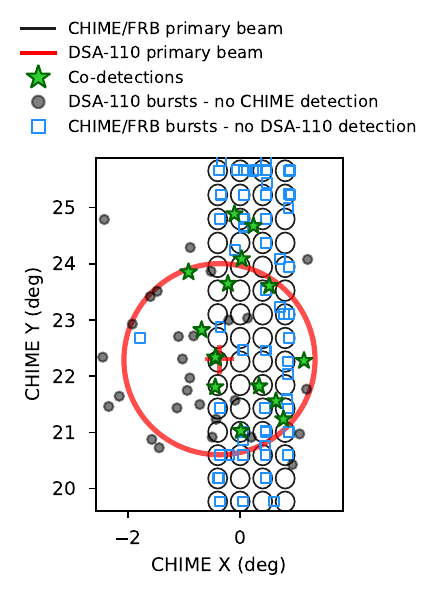}
    \caption{Overlap between the DSA-110 field-of-view and the subset of CHIME/FRB formed beams surrounding the DSA-110 pointing ($\sim5$\% of the total CHIME/FRB field of view), shown in CHIME/FRB beam model coordinates (X: E-W, Y: N-S, both in degrees from zenith). Black ellipses show the full width at half maximum (FWHM) of the CHIME/FRB formed beams at 568\,MHz \citep{chimecat1}. The red circle shows the DSA-110 primary beam (FWHM $3.4^\circ$) for its pointing at $\delta = 71.6^\circ$. Green stars mark the positions of the $15$ codetections presented in this work; grey circles mark DSA-110 FRBs detected between February 2022 and February 2024 without a CHIME/FRB counterpart (pulled from the DSA-110 event archive: \url{https://code.deepsynoptic.org/dsa110-archive/}), and blue squares mark CHIME/FRB Catalog 2 bursts from this same time period without a DSA-110 counterpart \citep{chimecat2}.}
    \label{fig:sky_overlap}
\end{figure}

The FRBs presented in this work were co-detected by both telescopes during the DSA-110 commissioning run from 2022 to 2024, prior to completion of the full DSA-110 array. During this time, both DSA-110 and CHIME/FRB were running near-continuously (barring technical interruptions), DSA-110 at a fixed declination of $\delta = 71.6^\circ$ (beam full-width-half-maximum of $3.4^\circ$) and CHIME over a declination range $-10^\circ < \delta < 90^\circ$. Based on their respective fields-of-view ($\approx10$ deg$^2$ for DSA-110 and $\approx200$ deg$^2$ for CHIME), we estimate that both telescopes were observing the same sky for $<5\%$ of the time period over which we search for co-detected FRBs \footnote{This estimated upper limit is only based on the field-of-view and ignores exposure times and sidelobe detections.}. Figure\,\ref{fig:sky_overlap} shows the overlap between the two telescopes' fields-of-view.

The DSA-110 commissioning survey operated over $1.28-1.53$\,GHz (optimal bandpass $1.3-1.5$ GHz) with a real-time FRB search at time and frequency resolutions of 262 $\mu$s and 244 kHz, respectively, up to a maximum DM of 1500 pc cm$^{-3}$. FRB candidates were triggered based on three criteria: (i) detection S/N\,$>$\,$8.5$ in any beam, (ii) no association with strong terrestrial radio frequency interference (RFI), and (iii) $\DM>50$\,pc\,cm$^{-3}$ and $\DM>0.75\times\DM_{\rm NE2001}$, the Galactic disk contribution to DM based on \cite{ne20011}. Subsequent to FRB candidate triggers, voltage data were recorded at time and frequency resolutions of 32.7\,$\mu$s and $30.5$\,kHz, respectively. For the nominal system equivalent flux density (SEFD) of 140 Jy at boresight, the fluence threshold of the real-time search was approximately 1 Jy ms (assuming a $\delta$-function pulse at a time resolution of 262 $\mu$s). 

The CHIME/FRB real-time search operates from $400-800$\,MHz at approximate time and frequency resolutions of 0.983\,ms and 24.4\,kHz, respectively, up to a maximum DM of 13,000\,pc\,cm$^{-3}$ \citep{chimefrb_overview}. FRB candidates were triggered if they exceeded a detection S/N of $8$, with a threshold of $\rm S/N>12$ required to trigger writing of the full-polarization voltage data with approximate time and frequency resolutions of 2.56\,$\mu$s and $390$\,kHz, respectively. Candidates are classified as FRBs if their detection DM exceeds the maximum predicted by both Galactic disk electron density models NE2001 \citep{ne20011} and YMW16 \citep{ymw16}. The median fluence threshold of the real-time search ($95\%$ completeness) is about 3.5\,Jy\,ms \citep{chimecat2}.

\subsection{Identifying Co-detections} \label{sec:identify_codetect}

\begin{figure*}
    \centering
    \includegraphics[width=0.95\textwidth]{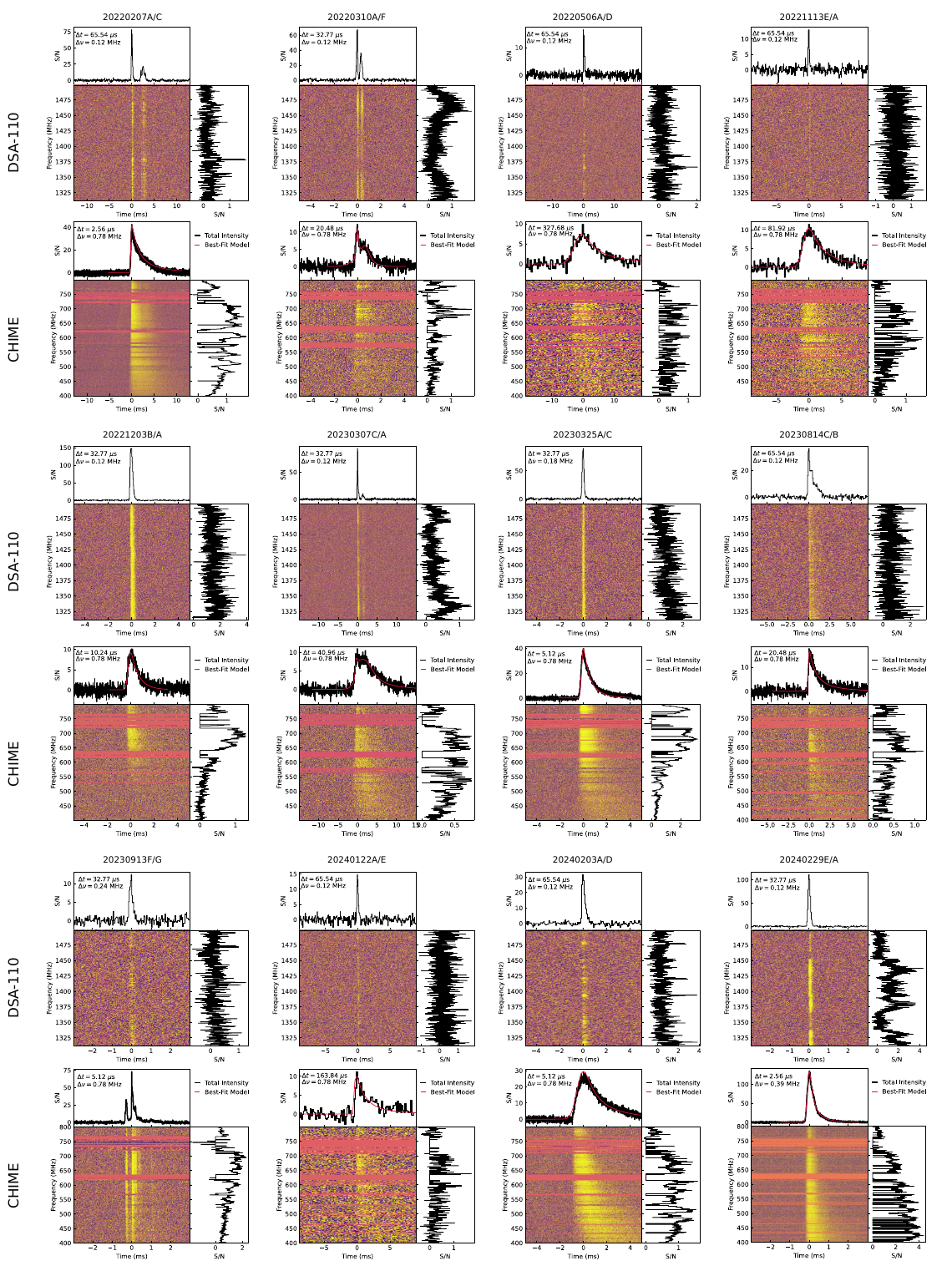}
    \caption{Dynamic spectra of 12 FRBs co-detected by DSA-110 and CHIME/FRB. Frequency channels dominated by RFI are masked and set to the mean off-burst noise level. Frequency-averaged burst intensity vs. time is shown in the top panels and time-averaged burst intensity vs. frequency in the righthand panels, both in units of peak S/N. Best-fit models of the CHIME/FRB burst profiles are shown in red (see Section~\ref{sec:pulse-broadening}). Time is shown relative to the burst peak in each band and does not indicate the absolute arrival time.}
    \label{fig:all_ds}
\end{figure*}

We begin with the DSA-110 FRB sample and ask whether CHIME/FRB detected a burst from the same source at a consistent DM, sky position, and time-of-arrival (ToA). We define a match according to the following three criteria: 
\begin{enumerate}
    \item The DMs inferred by the respective search pipelines are consistent to within 0.5 pc cm$^{-3}$, comparable to the root-mean-square difference between DMs measured from CHIME/FRB intensity and baseband data \citep{chime_basecat1}. 
    \item The sky positions are consistent to within $3\sigma$ between the two surveys. The burst localizations recorded at detection are used to generate an initial candidate list. Since the CHIME/FRB intensity data localization precision is $\approx700\times$ larger than the DSA-110 localization precision of $\approx 1^{\prime\prime}$, we subsequently check the candidate co-detections using refined CHIME/FRB localizations derived from the second CHIME baseband catalog (CHIME/FRB Collaboration in prep.); further details are provided later in this subsection.
    \item The difference in burst ToA recorded in real-time at the two telescopes, accounting for DM errors and burst widths, is $\Delta t_{\rm ToA} \leq \Delta t_{\DM} + \Delta t_{\rm c}$, where $\Delta t_{\DM}$ is the time delay expected from dispersion between the lowest frequency observed by DSA-110 (1.28 GHz) and the highest frequency observed by CHIME/FRB (800 MHz) and $\Delta t_{\rm c}$ is the light travel-time between the two observatories. 
\end{enumerate}

Comparing the DSA-110 FRB sample against CHIME/FRB real-time extragalactic triggers using the criteria above yields {14} candidates.\footnote{FRB~20220521B was detected by CHIME/FRB and just below the DSA-110 threshold, with consistent DMs but ToAs differing by tens of seconds. Given a chance-coincidence probability of $\lesssim10^{-6}$, these are likely separate bursts from the same repeating source. We exclude it, since it fails our selection criteria and no DSA-110 voltage data were saved.}. An additional burst, FRB~20220319D \citep{ravi2025}, was identified as a codetected source separately: CHIME/FRB detected it (DM$=111.2\pm0.4$\,pc\,cm$^{-3}$, S/N$=47$, RA, Dec $=$ 33.65$^\circ$, $+71.00^\circ$), but because its DM lies below the maximum Milky Way contribution along its LOS predicted by NE2001 ($\approx139$\,pc\,cm$^{-3}$; \citealt{ne20011}) and YMW16 ($\approx209$\,pc\,cm$^{-3}$; \citealt{ymw16}), it was classified as Galactic in real time and was not issued as an FRB trigger. Nonetheless, it satisfies all three criteria, and we include it for a total of {15} codetections. Since 20220319D was discovered through a different search, it breaks the uniformity of the sample, however, as no CHIME/FRB intensity or baseband data were saved, it is excluded from the analysis presented in this work. Requiring the bursts to have saved voltage data from both telescopes results in the final sample of 12 bursts presented below; their dynamic spectra are shown in Figure~\ref{fig:all_ds}, where the CHIME/FRB data are beamformed using methods described in \citet{Michilli_2021_ApJ}. Of the three co-detected FRBs without saved voltage data, only one has saved total intensity data from CHIME/FRB\footnote{For FRB~20220726A, only intensity data were saved, as its emission, concentrated at the top of the CHIME band, arrived before the baseband buffer began. FRB~20230712A was captured in neither intensity nor baseband data, owing to an unidentified technical fault (possibly latency). FRB~20220319D was classified as Galactic as described earlier.}; its dynamic spectrum is shown in Appendix~\ref{app:extra-codetections}. In Paper II, we show that bursts with detectable RMs in both the CHIME/FRB and DSA-110 data are entirely self-consistent, affirming that the selection criteria above are successful in identifying co-detected bursts.

One of the FRBs in the sample, 20220207A/C, has a $>3\sigma$ discrepancy between the CHIME/FRB localization from baseband data and the DSA-110 localization. CHIME/FRB baseband data give an (RA, Dec; J2000) of 20:40:36.16   $+$72:54:32.07 ($\pm3.84$s, 
$16.2^{\prime\prime}$), whereas DSA-110 data give 20:40:47.886 $+$72:52:56.378  ($\pm0.042$s, 
$0.54^{\prime\prime}$), amounting to an $\approx100^{\prime\prime}$ offset between the two reported positions. A series of cross-checks for each telescope's localization are described in Appendix~\ref{app:zach}. Given that the two telescopes observe consistent DMs and RMs, and the difference between the barycentric ToAs is $\approx0.6$ ms, we are confident that the source is indeed a co-detected FRB. We adopt the DSA-110 localization and host association of \citet{law2024}, but caution that a different true position could change the host association probability and any host-dependent quantities (e.g., redshift, $\rm DM_{host}$, $\rm RM_{host}$) derived for this FRB.

\begin{figure}
    \centering
    \includegraphics[width=0.49\textwidth]{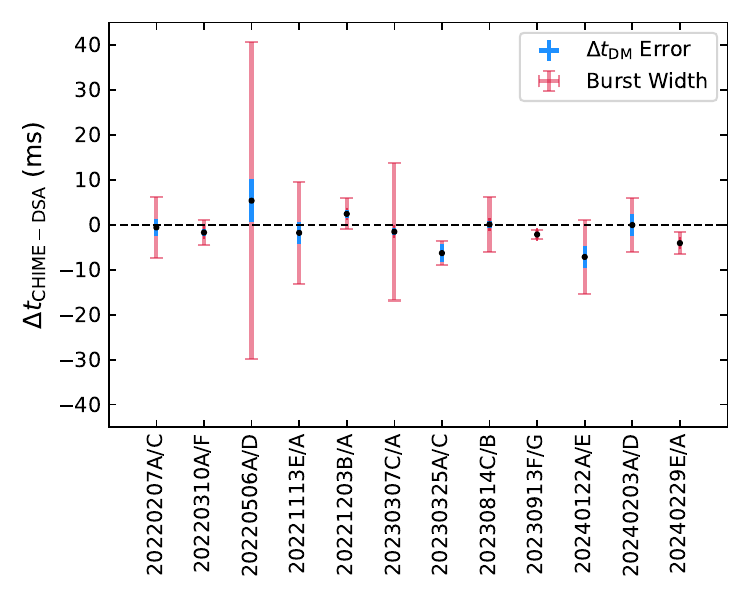}
    \caption{Difference in barycentric time-of-arrival (ToA) between CHIME/FRB and DSA-110, after correcting for dispersion, for the twelve co-detected FRBs with saved voltage data. Error bars indicate contributions to the ToA uncertainty from the uncertainty in the burst DM (blue) and the total burst width (red; conservatively shown here as the sum of burst widths measured at CHIME/FRB and DSA-110, although the CHIME/FRB widths dominate in all cases; see Section~\ref{sec:spectra}). 
    }
    \label{fig:toa-diffs}
\end{figure}

We note that, for each of the $15$ co-detected bursts, only one CHIME/FRB detection on the same day satisfies the DM and position criteria (1)--(2), before the ToA criterion (3) is even applied. Furthermore, none of the DSA-110-detected FRBs have been associated with repeating FRBs detected by CHIME/FRB. Nonetheless, these selection criteria are agnostic to whether the emission detected by both CHIME/FRB and DSA-110 represents the same burst component(s) spanning $0.4-1.5$ GHz, or whether the emission arises from multiple burst components, some of which are seen by one telescope and not the other. Figure~\ref{fig:toa-diffs} shows the difference in barycentric ToA between the two telescopes after correcting for dispersion; nearly all of the FRBs have ToA differences that are consistent with zero after considering uncertainties due to the burst width and DM (which are somewhat degenerate with each other, but shown separately in Figure~\ref{fig:toa-diffs} for illustrative purposes). The burst ToA difference between the two observatories is not corrected for any potential clock errors or systematic offsets between the clocks at each observatory; pulsar timing experiments indicate that such errors can occasionally be as large as milliseconds \citep[e.g.][]{lazarus2016}. Studies of highly active repeating FRBs also indicate that the waiting time between bursts can be as small as a few milliseconds \citep{li2021}. Given the range of $\Delta t_{\rm ToA}$ inferred is between $0.05-7$ ms (excluding uncertainties), we thus cannot make robust conclusions about the correspondence between burst components seen by each telescope. In Section~\ref{sec:spectra} we assess additional evidence for and against the observed emission representing multiple independent burst components.

The FRB surveys conducted by CHIME and DSA-110 are subject to their own unique systematics, which undoubtedly affect the number of co-detected FRBs we identify and their properties. DSA-110 is optimally sensitive to temporal burst widths between $0.26-8.32$ ms \citep{ravi2023,law2024}, while CHIME/FRB is optimally sensitive to temporal burst widths between $0.1-10$ ms (although it searches at a temporal resolution of 1 ms; \citealt{chimecat1,mcgregor2026, merryfield2023}). A slew of other systematics affect burst detection rates, including the unique RFI environments of each telescope, their sensitivities, etc. A comprehensive analysis of how these systematics affect the rate of co-detected bursts is beyond the scope of this study; a detailed assessment of CHIME/FRB systematics can be found in \cite{chimecat1} and \cite{mcgregor2026}, and a similar assessment for the DSA-110 is the subject of future work. In Section~\ref{sec:selection}, we consider further the implications of comparing burst properties between the co-detected FRBs and the broader CHIME/FRB and DSA-110 samples.

\subsection{Sample Overview}\label{sec:sample-overview}

\begin{deluxetable*}{c c c c c c c c c}
\label{tab:burst-props}
\tablecaption{Co-detection Sample Properties}
\tabletypesize{\footnotesize}
\tablehead{\colhead{TNS Name} & {Localization (J2000)} & Error (arcsec) & {DM (pc cm$^{-3}$)} & \colhead{$z$} & {\DM$_{\rm host}$ (pc cm$^{-3}$)} & \colhead{$\tau$ (ms at 1 GHz)} & \colhead{$\Delta \nu_{\rm d}$ (MHz at 1 GHz)} & \colhead{$\Delta \nu_{\rm d}$ index}}
\startdata
20220207A/C & 20:40:47.886 $+$72:52:56.378 & (0.63,0.54) & $262.433\pm0.004$ & 0.0430 & $140^{+19}_{-27}$ & $0.2910\pm0.0004$ & $0.24\pm0.03$ & $3.7\pm0.3$\\ 
20220310A/F & 08:58:52.92 $+$73:29:27.0 & (1.9,1.4) & $462.21\pm0.03$ & 0.4779 & $137^{+68}_{-79}$ & $0.01737\pm0.00002$ & $5.4\pm1.8$ & $4.4\pm0.8$\\ 
20220506A/D & 21:12:10.760 $+$72:49:38.20 & (1.11,0.81) & $397.11\pm0.05$ & 0.3004 & $133^{+50}_{-70}$ & $0.65\pm0.03$ & $0.88\pm0.14$ & $3.9\pm0.4$ \\ 
20221113E/A & 04:45:38.64 $+$70:18:26.6 & (1.4,0.9) & $411.70\pm0.02$ & 0.2505  & $173^{+48}_{-76}$ & $0.327\pm0.009$& - & -\\ 
20221203B/A & 21:00:31.09 $+$72:02:15.22 & (0.46,0.56) & $602.34\pm0.04$ & \nodata & \nodata & $0.139\pm0.003$& $0.10\pm0.02$ & $3.9\pm0.5$ \\ 
20230307C/A & 11:51:07.52 $+$71:41:44.3 & (0.2,0.2) & $610.577\pm0.005$ & 0.2710  & $463^{+57}_{-127}$ & $0.237\pm0.005$ & $1.5\pm0.6$ & $5.0\pm0.9$\\ 
20230325A/C & 05:52:45.43 $+$74:12:00.37 & (1.6,1.0) & $912.469\pm0.001$ & \nodata & \nodata & $0.1299\pm0.0005$ & $0.21\pm0.05$ & $4.2\pm0.6$ \\ 
20230814C/B$^*$ & 22:23:53.94 $+$73:01:33.26 & (1.7,0.5) & $696.50\pm0.03$ & 0.5535 & $310^{+103}_{-154}$ & $0.151\pm0.003$ & $0.17\pm0.03$ & $4.4\pm0.5$ \\ 
20230913F/G & 20:20:08.92 $+$70:47:33.96 & (0.8,0.5) & $518.796\pm0.002$ & 0.3024 & $286^{+59}_{-111}$ & $0.002722\pm0.000006$& - & -\\ 
20240122A/E & 02:39:03.96 $+$71:01:04.3 & (1.3,0.7) & $960.20\pm0.02$ & \nodata & \nodata & $0.23\pm0.01$& $0.56\pm0.12$ & $5.1\pm0.4$ \\ 
20240203A/D & 20:50:28.59 $+$73:54:00.0 & (0.8,0.6) & $272.6\pm0.1$ & 0.0740 & $134^{+23}_{-34}$ & $0.1475\pm0.0009$ & $0.40\pm0.04$ & $3.8\pm0.2$ \\ 
20240229E/A & 11:19:56.05 $+$70:40:34.4 & (0.8,0.6)  & $491.207\pm0.002$ & 0.2870 & $305^{+56}_{-112}$ & $0.01945\pm0.00002$& - & -\\ 
\hline
20220319D$^\dagger$ & 02:08:42.7 $+$71:02:06.9 & (0.58,0.55) & $111$ & 50 Mpc & & & \\ 
20220726B/A & 05:00:25.389 $+$71:36:06.54 & (0.8,0.6) &  $686$ & 0.3610 & & & \\ 
20230712H/A & 11:22:50.688 +71:46:45.75 & (1.4,0.9) & $587$ & 0.4525 & & & \\ 
\enddata
\tablecomments{Left to right: Source name (Transient Name Server, TNS, convention with CHIME/FRB/DSA-110 event letters listed side-by-side); localization reported by the DSA-110 \citep{law2024,sharma2024,verdi2026}; dispersion measure (DM) based on fits to the CHIME/FRB data (see Section~\ref{sec:sample-overview}); host galaxy redshift (from \citealt{law2024,connor2025,verdi2026}); host galaxy contribution to DM (see Section~\ref{sec:screens}); pulse broadening time at 1 GHz (Section~\ref{sec:pulse-broadening}), where the quoted uncertainties are statistical uncertainties from the fit and do not account for uncertainty in the frequency scaling ($\tau\propto\nu^{-4}$) assumed to convert to 1\,GHz; scintillation bandwidth at 1 GHz; and the frequency-scaling index of the measured scintillation (Section~\ref{sec:scintillation}). Two co-detected bursts lacking sufficient S/N and baseband voltage data for subsequent analysis are listed at the bottom. $^*$20230814C/B has been observed to repeat \citep{johndoe_atel}. $^\dagger$20220319D was not reported to TNS by CHIME (see text).} 
\end{deluxetable*}

Table~\ref{tab:burst-props} presents an overview of the co-detected FRB sample and key properties. Polarization fractions and RMs are separately presented in Paper II. All but one source in the sample are as-yet non-repeating; FRB 20230814C/B was observed to repeat once by the DSA-110 \citep{johndoe_atel}.

The sample covers a wide range in observed DM ($262-960$ pc cm$^{-3}$). We measure the DM of each FRB independently in the CHIME/FRB and DSA-110 bands\footnote{The uncertainty in the burst ToAs at each telescope precludes a joint fit of the data across the full $0.4-1.5$ GHz frequency coverage.}, verifying that the two measurements are consistent. As the bursts in our sample predominantly have simple morphologies, we determine the DM by maximizing the peak flux within each band, which for such bursts converges with structure-optimizing approaches \citep[e.g.][]{hessels2019}; we have visually verified that all bursts appear correctly dedispersed at the adopted DMs. The reported DM uncertainty is based on the DM at which the peak flux drops by $5\%$ \citep[e.g.][]{wang2026}. For bursts with significant pulse broadening in the CHIME/FRB band, the peak-flux-maximizing DM is biased high; so for these bursts we additionally require that the leading edge of the FRB be aligned across radio frequency. We find that the independently measured DMs are consistent between the two bands, generally agreeing to within $2\sigma$. Since CHIME/FRB's larger frequency bandwidth and lower central frequency yields significantly more precise DM measurements, we quote CHIME/FRB-measured DMs in Table~\ref{tab:burst-props}.

Nine of the FRBs were localized by DSA-110 to host galaxies with secure redshifts (independently reported in \citealt{law2024,sharma2024,connor2025,verdi2026}). The redshifts are shown in Table~\ref{tab:burst-props}. These galaxies are all at $z\lesssim0.6$; the two highest-DM FRBs in the sample (20230325A/C and 20240122A/E, both at $\DM>900$ pc cm$^{-3}$) currently lack host galaxy associations likely due to insufficiently deep optical/infrared imaging (although even with deep imaging, the localization regions may not be precise enough to associate each FRB with a single galaxy). FRB 20221203B/A requires additional spectroscopic characterization for redshift determination.

\section{Morphology and Energetics}\label{sec:spectra}

FRBs are often classified by their spectral morphologies, e.g., number of components, frequency-time drift, spectral occupancy, spectral index, and width \citep{hessels2019,macquart2019,pleunis2021,scott2025}. Compared to previous studies of both repeating and non-repeating FRBs, which generally find a diverse range of spectral morphologies \citep[e.g.][]{zhou2022,faber2024_chime,scott2025,curtin2025}, the co-detected sample presented here is somewhat uniform. Nearly all of the co-detected FRBs show emission extending across the entire frequency coverage of both telescopes. Only a few of the FRBs show a significant decrease in peak flux density below 600\,MHz accompanied by significant pulse broadening, although the source spectral index may also play a role (see Section\,\ref{sec:spectral-index}). We characterize the morphologies of these bursts first by comparing their shapes and number of components between the two frequency bands, followed by a comparison of approximate burst fluences, flux densities, and energies.

\subsection{Burst Shapes}
Half of the sample shows a single component in both the DSA-110 and CHIME/FRB data. Three FRBs (20220207A/C, 20220310A/F and 2030307C/A) show two prominent components at $1.28-1.53$\,GHz that appear blended by scattering at $400-800$\,MHz (Figure\,\ref{fig:all_ds}). The repeating FRB~20230814C/B is likewise multi-component in the DSA-110 data but a single scattered component in the CHIME/FRB data. All FRBs except FRB~20230913F/G show visibly significant temporal broadening at $400-800$\,MHz (Section~\ref{sec:pulse-broadening}). FRB~20230913F/G is instead the only burst with clear multi-component structure in the CHIME/FRB data, yet it shows a single component at $1.28-1.53$\,GHz.

ToA uncertainties and the telescopes’ differing sensitivities and time resolutions (32.76\,$\mu$s for DSA-110 vs. 2.56\,$\mu$s for CHIME/FRB) prevent a one-to-one matching of components, so the blending interpretation above rests on morphology alone. No FRB, including the repeater, shows robust frequency–time drift, a common hallmark of repeating FRBs \citep{pleunis2021}, although scattering may hide multi-component structure in the CHIME/FRB data.

\subsection{Fluences}\label{sec:fluence-flux}

Owing to differences in telescope sensitivity we estimate the fluences and mean flux densities for each burst, rather than directly comparing the burst S/N between the DSA-110 and CHIME/FRB data.  
For the DSA-110 data, the fluence is calculated using the frequency-averaged burst profile, converted from S/N units to Jy and then summed over its $99\%$ temporal width at the smallest time resolution achievable, $32.7\ \mu$s. We estimate the effective SEFD at the location of the FRB within the primary beam assuming a Gaussian beam response model, with an SEFD at boresight of 140 Jy. The Gaussian beam response model is an approximation of the true beam response, which is not well-characterized, and  
in subsequent analysis we thus adopt a conservative fluence uncertainty of $\pm30\%$. 

The larger frequency bandwidth of CHIME allows us to examine both the average fluence across the entire 400-800\,MHz band, as well as the fluence as a function of frequency. When the CHIME data are beamformed \citep{Michilli_2021_ApJ}, the amplitudes of the total intensity data are calibrated with daily gains solutions. The data are then downsampled in time to improve the S/N for on-burst flux determination. We subtract the average off-burst flux density to correct any differences between the calibration and FRB arising from, e.g., changes in the sky temperature, and convert to Jy units using a beamforming factor that corrects for the coherent gain of the tied-array sum. We then correct for the response of the primary beam \citep{Amiri_2023_ApJ}. \citet{chime_basecat1} perform this correction using the average beam response, whereas in this work we divide out the frequency-dependent response to explore the average fluence {\it and} the fluence spectra of the bursts. The average fluence per burst is then estimated by summing over frequency channels, polarization hands, and the time window that encapsulates 99\% of the total flux.

To examine the fluence as a function of frequency, we consider two methods for defining the on-burst interval. First, we use the 99\% width that we use above. However, in the presence of a frequency-dependent burst width (e.g., the pulse broadening seen in most of the CHIME bursts in this sample; Figure\,\ref{fig:all_ds}), a single, frequency-independent window can exclude a substantial fraction of the burst flux at lower frequencies, biasing the measured spectral index. We therefore also consider an alternative burst-extent determination using {\tt fitburst} modeling \citep{fonseca2024} of the FRB dynamic spectra (see Section~\ref{sec:scattering-analysis} for details) to define the on-burst region independently per frequency channel: bins where the {\tt fitburst} model exceeds 20\% of that channel's peak model amplitude are included as ``on-burst" with all other bins set to zero. This approach accounts for the frequency-dependent burst duration when calculating the fluence spectrum. For FRB~20230913F/G, whose complex burst structure prevented a reliable {\tt fitburst} fit, we instead use only the 99\%-width method for this burst; we do not expect this choice to introduce a significant bias in subsequent analysis, since this FRB has the smallest measured scattering in our sample and is therefore least affected by a frequency-dependent burst width.

\begin{figure*}
    \centering
    \includegraphics[width=0.95\textwidth]{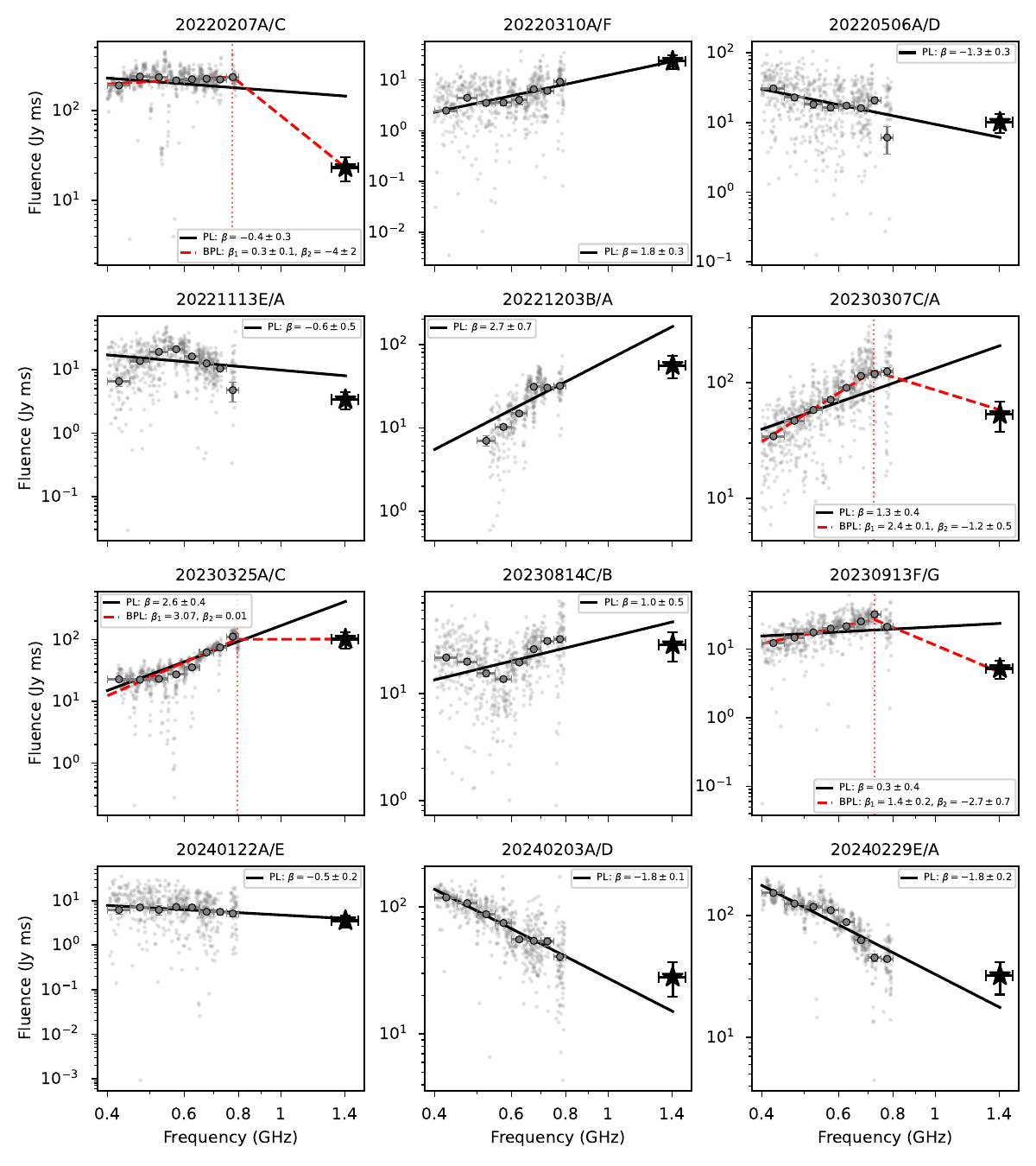}
    \caption{Fluence vs. frequency for the $12$ bursts in our sample, integrated over the {\tt fitburst} frequency-dependent temporal widths defined in Section\,\ref{sec:fluence-flux}. Faint grey points show the per-channel CHIME fluence; the larger grey points with error bars show the CHIME fluence averaged in eight 50\,MHz subbands, with uncertainties given by the root-mean-square of the channel-wise fluences within the subband normalized by the square-root of the number of channels. The black star marks the DSA-110 fluence calculated over the entire DSA-110 band. The black solid line is a single power-law (PL) fit to both CHIME and DSA-110 measurements, with a best-fit spectral index $\beta$ labeled in each panel; quoted uncertainties are rescaled by $\sqrt{\bar{\chi}^2}$ for fits with reduced chi-square $\bar{\chi}^2>1$. Four FRBs (20220207A/C, 20230307C/A, 20230325A/C, and 20230913F/G) show evidence of a spectral turnover (see Section~\ref{sec:spectral-index}).
    For these cases, we show a broken power-law fit (BPL; red dashed line) as a convenient parametrization of the turnover, with the rising and declining spectral indices $\beta_1$ and $\beta_2$ labeled, and the fitted break frequency shown by the vertical red dotted line. In all of these cases the spectrum appears to peak near the upper edge of the CHIME frequency band, and these fits should thus be treated with caution.}
    \label{fig:ind_spec}
\end{figure*}

\subsection{Spectral Index}\label{sec:spectral-index}

We explore constraints on the FRB spectral index by combining the fluence spectra across the CHIME/FRB band with 
the average fluences inferred from the DSA-110 data.  
We do not explore the evolution of the DSA-110 fluences across frequency due to the lack of absolute flux calibration. The resulting fluence spectra and spectral index fits (using the {\tt fitburst}-widths defined in Section\,\ref{sec:fluence-flux}) are shown in Figure~\ref{fig:ind_spec}, with the spectral indices reported in Table~\ref{tab:spec_props}. We caution that this sample is biased (see Section\,\ref{sec:selection}): e.g. co-detection requires the burst to be bright enough for both CHIME/FRB and DSA-110, which may favor spectra that are either flat or peak between the two frequency bands. We find no significant correlation between the scattering timescale (Table\,\ref{tab:burst-props}) and the spectral index (Spearman $\rho=-0.31$, $p=0.33$), suggesting that the spectral indices are not biased by scattering.

\begin{figure}
    \centering
    \includegraphics[width=0.47\textwidth]{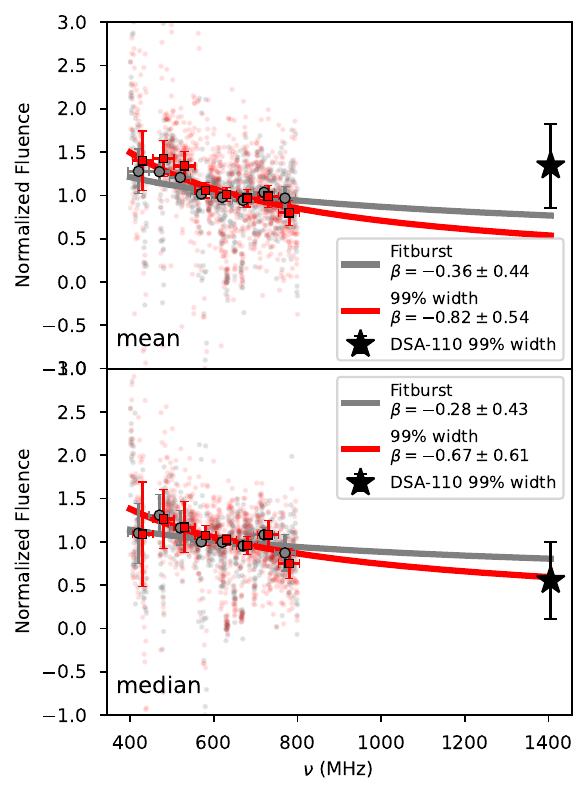}
    \caption{Combined fluence spectra for the 12 bursts, computed as the mean (top panel) and median (bottom panel) of each burst's fluence spectrum after normalizing by its median fluence within the CHIME band. For the CHIME data (400--800\,MHz), the faint points show the channel-wise combined spectrum and the larger solid points show the same spectrum averaged in eight 50\,MHz subbands, with uncertainties based on standard error for the mean in the top panel and bootstrap resampling over bursts in the bottom panel. The grey points and line show the spectrum and power-law fit obtained using the {\tt fitburst}-determined, frequency-dependent burst extent; the red points and line correspond to using a fixed (frequency-independent) window encapsulating 99\% of the burst fluence. The black star at 1405\,MHz is the combined DSA-110 data point, where each individual spectrum is extracted using a 99\%-width window and normalized by the same per-burst CHIME normalization factor, with uncertainties computed in the same way as in the CHIME data. In each panel, the quoted spectral index $\beta$ is from a power-law fit to both the CHIME and DSA-110 fluences, with uncertainty from bootstrap resampling over bursts.}
    \label{fig:comb_spec}
\end{figure}

Following \citet{macquart2019}, we combine the individual spectra into a sample-mean and sample-median fluence spectrum, shown in Figure~\ref{fig:comb_spec}. To prevent any single burst from dominating the combined spectrum, we normalize each burst's fluence spectrum by its own median before combining the sample. Using the {\tt fitburst}-defined on-burst intervals, which account for frequency-dependent burst durations, and which we therefore consider the more reliable method, we find sample-mean and sample-median spectral indices of $\beta = -0.4\pm0.4$ and $\beta = -0.3\pm0.4$, respectively. The frequency-independent 99\% width yields steeper values (although consistent within uncertainties) of $\beta = -0.8\pm0.5$ and $\beta = -0.7\pm0.6$ for the sample-mean and median, respectively, consistent with previous measurements \citep{macquart2019, shin2022}.

Four bursts in our sample (FRBs 20220207A/C, 20230307C/A, 20230325A/C, and 20230913F/G) show evidence for a spectral turnover: the DSA-110 fluence lies significantly below the power law extrapolated from the CHIME/FRB band alone, in some cases by more than an order of magnitude, and a single power law provides a poor description of the data ($\bar{\chi}^2\approx37$--$86$). For all four the CHIME/FRB spectrum rises to the top of the band, yet the DSA-110 fluence is lower than the CHIME/FRB fluence at 800\,MHz, implying that the spectrum reaches a maximum somewhere between the two bands. To illustrate the spectral shape we fit a broken power law, $S(\nu)\propto (\nu/\nu_{\rm peak})^{\beta_1}$ below $\nu_{\rm peak}$ and $S(\nu)\propto (\nu/\nu_{\rm peak})^{\beta_2}$ above (Figure\,\ref{fig:ind_spec}), with $\beta_1$ fixed to a CHIME-only power-law fit and only $\beta_2$ and $\nu_{\rm peak}$ left as free parameters. The broken power law is strongly preferred over a single power law for FRBs 20220207A/C, 20230307C/A, and 20230913F/G (F-test $p\approx8\times10^{-5}$--$2\times10^{-3}$), and provides an acceptable fit for FRBs 20230307C/A and 20230913F/G ($\bar{\chi}^2\approx1.2$ and $1.7$), but not for FRB~20220207A/C ($\bar{\chi}^2\approx10$), whose fitted break lies near the upper edge of the CHIME band. For FRB~20230325A/C, the DSA-110 fluence lies below the extrapolated CHIME-only power law, but neither model describes the spectrum well ($\bar{\chi}^2\approx64$ and $59$), the broken power law is not statistically preferred (F-test $p\approx0.35$), and its break frequency and declining index are unconstrained. We emphasize that the broken power law is a convenient parametrization rather than a physical model: the data do not distinguish a sharp break from a smoothly curved spectrum. Since only a single data point lies above the break, the break frequency and the high-frequency spectral index are poorly constrained. We report the fit results in Table\,\ref{tab:spec_props}, but these constraints should be treated with caution; see Section~\ref{sec:disc-scatter} for further discussion.

If spectral breaks are a generic property of FRBs, the single power-law fits in Table~\ref{tab:spec_props} constrain the break location: bursts with positive (negative) indices $\beta$ must peak above (below) our observing band. Under this assumption, $4/12$ bursts peak within 400--1500\,MHz, $5/12$ peak below 400\,MHz, and $3/12$ peak above 1.5\,GHz. The sample median break frequency is $\nu_{\rm peak} \approx 720$\,MHz, while the full distribution may span at least a factor of $4$ in frequency. FRB spectral peaks therefore do not cluster at a single characteristic frequency. 

\begin{deluxetable*}{c c c c c c}
\label{tab:spec_props}
\tablecaption{Spectral fitting results}
\tabletypesize{\footnotesize}
\tablehead{\colhead{TNS Name} & \colhead{PL $\beta$ \tablenotemark{a}} & \colhead{BPL $\beta_1$ \tablenotemark{b}} & \colhead{BPL $\beta_2$ \tablenotemark{b}} & \colhead{$\nu_{\rm peak}$ (MHz) \tablenotemark{c}} & \colhead{$E_{400-1500\,{\rm MHz}}$ ($10^{40}$\,erg) \tablenotemark{d}} }
\startdata
20220207A/C & $-0.4\pm0.3$ & $0.28\pm0.12$ & $-3.9\pm1.6$ & $774\pm31$ & $0.60\pm0.12$\\ 
20220310A/F & $1.8\pm0.3$ & \nodata & \nodata & $>1500$ & $8.4\pm1.7$ \\ 
20220506A/D & $-1.3\pm0.3$ & \nodata & \nodata & $<400$ & $3.2\pm0.4$ \\ 
20221113E/A & $-0.6\pm0.5$ & \nodata & \nodata & $<400$ & $1.9\pm0.4$ \\ 
20221203B/A & $2.7\pm0.7$ & \nodata & \nodata & $>1500$ & \nodata \\ 
20230307C/A & $1.3\pm0.4$ & $2.39\pm0.12$ & $-1.2\pm0.5$ & $721\pm14$ & $17\pm2$\\ 
20230325A/C & $2.6\pm0.4$ & $3.1\pm0.4$ & $0.0$\tablenotemark{e} & $794$\tablenotemark{e} & \nodata \\ 
20230814C/B & $1.0\pm0.5$ & \nodata & \nodata & $>1500$ & $29\pm9$\\ 
20230913F/G & $0.3\pm0.4$ & $1.37\pm0.18$ & $-2.7\pm0.7$ & $724\pm18$ & $3.5\pm0.4$\\ 
20240122A/E & $-0.5\pm0.2$ & \nodata & \nodata & $<400$ & \nodata\\ 
20240203A/D & $-1.76\pm0.12$ & \nodata & \nodata & $<400$ & $0.61\pm0.03$ \\ 
20240229E/A & $-1.8\pm0.2$ & \nodata & \nodata & $<400$ & $12.2\pm0.8$\\ 
\enddata
\tablenotetext{a}{Spectral index of a single power-law (PL) fit to the CHIME/FRB and DSA-110 fluences (Section~\ref{sec:spectral-index}).}
\tablenotetext{b}{Rising ($\beta_1$) and declining ($\beta_2$) indices of a broken power-law (BPL) fit, quoted only when the DSA-110 fluence lies $>3\sigma$ from the extrapolated CHIME-only PL fit. $\beta_1$ is fixed to the CHIME-only PL index.}
\tablenotetext{c}{Fitted BPL break frequency. For PL-only bursts, limits assume a break exists: positive (negative) $\beta$ implies $\nu_{\rm peak}>1500$\,MHz ($<400$\,MHz).}
\tablenotetext{d}{Isotropic-equivalent energy from integrating the fitted spectral model (BPL where a turnover is present, otherwise PL) over 400--1500\,MHz. Not given for the three bursts without redshifts.}
\tablenotetext{e}{BPL fit unconstrained; best-fit values without meaningful uncertainties.}
\tablecomments{Uncertainties are rescaled by $\sqrt{\bar{\chi}^2}$ for fits with $\bar{\chi}^2>1$ to account for unmodeled spectral structure.}
\end{deluxetable*}

\subsection{Burst Energy and Flux Density}\label{sec:energy}

We use the fluences to calculate spectral energies (in erg Hz$^{-1}$) for all of the FRBs with known redshifts. Following \cite{macquart2018}, we define the burst spectral energy as
\begin{equation}\label{eq:energy-fluence}
    E_\nu = \frac{4\pi D_L^2}{(1+z)^{2+\beta}}F_\nu
\end{equation}
where $D_L$ is the luminosity distance (calculated assuming the \citealt{planck18} cosmology), $z$ is the burst redshift, $\beta$ is the burst spectral index, and $F_\nu$ is the burst fluence. We evaluate Equation~\ref{eq:energy-fluence} for two different assumptions of the burst spectral index: $\beta=0$ (a flat spectral index), and $\beta$ given by fits to the fluence spectrum of each FRB (Table~\ref{tab:spec_props}), defined to be $S(\nu)\propto\nu^{\beta}$. For FRBs with evidence of a broken power-law spectrum, we use $\beta_1$ to evaluate spectral energy from the CHIME fluences and $\beta_2$ for the DSA-110 fluences.

\begin{figure}
    \centering
    \includegraphics[width=0.47\textwidth]{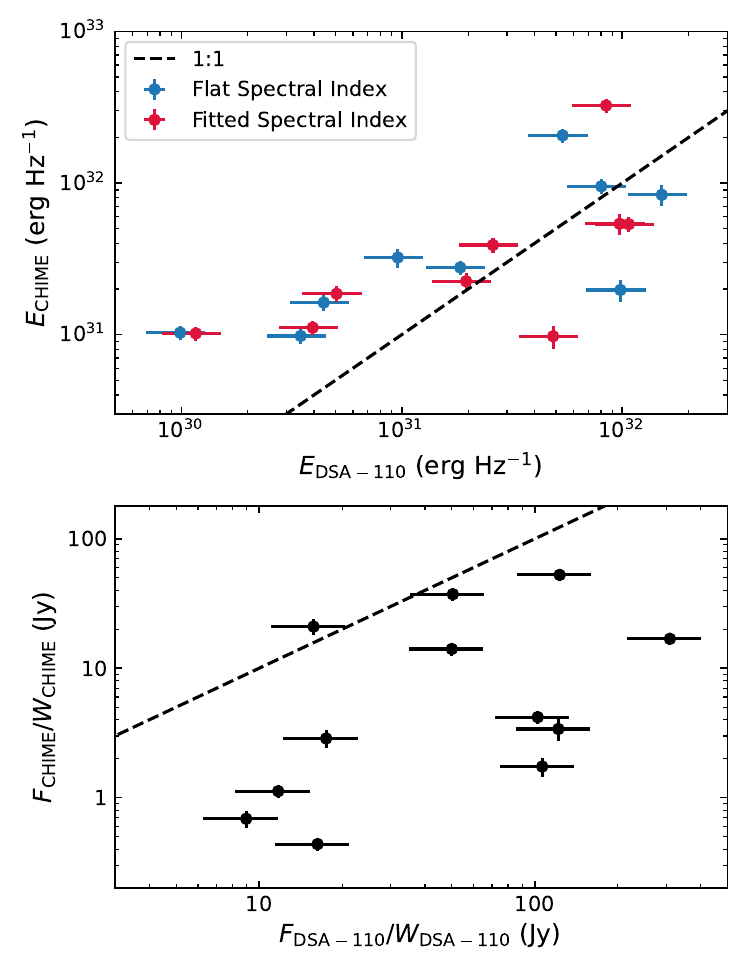}
    \caption{Top: Burst spectral energy $E$ estimated from CHIME fluences vs. energy estimated from DSA-110 fluences, for a flat spectral index (blue) and a fitted spectral index (red; see Section~\ref{sec:spectral-index}). The black dashed line indicates where $E_{\rm CHIME} = E_{\rm DSA-110}$. 
    Bottom: Mean flux density, taken to be fluence $F$ divided by burst width $W$, for CHIME vs. DSA-110. See Section~\ref{sec:fluence-flux} for details.
    }
    \label{fig:flux-comparison}
\end{figure}

Figure~\ref{fig:flux-comparison} compares the burst spectral energies estimated at 1.4\,GHz and 600\,MHz, using the DSA-110 and CHIME data, respectively. The CHIME spectral energies are larger than the DSA-110 spectral energies for all but 2-3 of the FRBs, depending on the assumed spectral index. Given that scattering typically conserves spectral energy (unless the scattering screen is truncated; \citealt{cl_trunc_screen2001}), a source with a flat spectral index should have a fluence and spectral energy that are nearly constant with frequency. The clear increase in spectral energy at $400-800$ MHz for most of the bursts is thus another indicator of the intrinsic spectral index. It is also likely that selection effects shape the distribution of codetected FRB spectral energies, since more highly scattered bursts would need to be more energetic to be detected. We explore the potential impact of such selection effects further in Section~\ref{sec:selection}.

We also compute the integrated energy of each burst across 400-1500\,MHz by integrating under the fitted spectral behavior presented in Figure\,\ref{fig:ind_spec} (where we use the broken power-law fits where applicable). We convert to an isotropic-equivalent energy, $E_{400-1500\,{\rm MHz}}$, using the luminosity distance calculated earlier. These energies are reported in Table\,\ref{tab:spec_props}, and are more representative of the ``total" energy than values determined from the CHIME or DSA-110 datasets alone, although we do not extrapolate beyond the CHIME and DSA-110 frequency bands. The uncertainties are calculated by Monte Carlo sampling each burst's power-law fit covariance (rescaled by $\sqrt{\bar{\chi}^2}$ with reduced chi-square $\bar{\chi}^2>1$) 2000 times and re-integrating.

Figure~\ref{fig:flux-comparison} also shows the mean flux density in each band, taken to be the fluence divided by the total burst width. The substantially larger burst widths at $400-800$\,MHz yield a significant difference in the median flux between the two bands: 3.8\,Jy for CHIME and 50\,Jy for DSA-110. Even after accounting for the large uncertainties in the DSA-110 estimates, it is clear that the FRBs are substantially higher in flux at $1.28-1.53$ GHz than at $400-800$ MHz, which is likely due to the decrease in scattering at higher frequencies.

\section{Scattering Analysis}\label{sec:scattering-analysis}

The broad frequency coverage of the co-detected sample provides an opportunity to search for both pulse broadening and scintillation. Pulse broadening produces a chromatic, temporally asymmetric increase in the observed burst width characterized by a scattering timescale $\tau$, while scintillation produces frequency-dependent modulations in the burst intensity that decorrelate over a characteristic scintillation bandwidth $\Delta\nu_{\rm d}$. Both effects are strongly frequency dependent, with $\tau \propto \nu^{-\alpha}$ and $\Delta \nu_{\rm d} \propto \nu^{+\alpha}$, making their signatures distinguishable from intrinsic burst structures when studied over a wide frequency range. For a Kolmogorov-like density spectrum with scattering well within the inertial range, $\alpha=4.4$; if scattering occurs on scales well below the inner scale, or the spectrum is much steeper than Kolmogorov, $\alpha=4$ \citep[e.g.][]{cordes1986,rickett1990}. The latter case, commonly assumed for FRBs, corresponds to a Gaussian scattered image and a one-sided exponential pulse broadening function (PBF). While real-world PBFs can deviate from this (owing to the inner scale, spectral index or truncated screens; \citealt{lambert99}), we assume a one-sided exponential which is sufficient for estimating characteristic scattering times and scintillation bandwidths. 

\subsection{Pulse Broadening}\label{sec:pulse-broadening}
We restrict our pulse broadening analysis to the CHIME/FRB baseband data, since scattering is strongly frequency dependent and its effects on the burst morphology are far more prominent at the lower observing frequencies of CHIME/FRB (Figure\,\ref{fig:all_ds}). An alternative approach would be to perform a joint fit to both the DSA-110 and CHIME data, but the DSA-110 data add little constraining power to the fit due to its higher central frequency and much narrower frequency coverage.
To measure the scattering timescale, we fit the Stokes I dynamic spectra of each burst using {\tt fitburst}, which assumes the burst consists of a set number of Gaussian components modified by an intrinsic spectral index, pulse broadening, and intrachannel dispersion smearing (\citealt{fonseca2024}, see also \citealt{curtin2025,sand2025}). In each fit, the DM is fixed to the value measured in Section\,\ref{sec:sample-overview} and the scattering time has a fixed frequency scaling of $\alpha=4$. The number of burst components included in each fit is chosen by visual inspection of the dynamic spectrum. For FRB~20230913F/G, we fit only the leading components of the burst to reduce the complexity of the fit, allowing for a more reliable scattering measurement. None of the FRBs have convergent fits when scattering is excluded from the burst model. The reduced chi-square statistic of the fits range between $\bar{\chi}^2\approx 0.88-0.99$ for most of the bursts; the fit for FRB 20240229E/A has the lowest $\bar{\chi}^2$ of $\approx 0.7$, indicating that the noise level is overestimated for this burst, rendering its uncertainties conservative. 
The best-fit scattering times, scaled to a reference frequency of 1\,GHz assuming $\tau\propto\nu^{-4}$, are reported in Table\,\ref{tab:burst-props} and shown in Figure~\ref{fig:all_ds}. 

\subsection{Scintillation}\label{sec:scintillation}

We search for evidence of diffractive scintillation in both the CHIME/FRB and DSA-110 observations by measuring the scintillation bandwidth from the autocorrelation function (ACF) of the time-averaged burst spectrum. 

For the DSA-110 observations, we use the highest frequency resolution data products available (0.0305\,MHz channels), which set the minimum measurable scintillation bandwidth. The burst dynamic spectrum is normalized by the off-burst standard deviation in each channel to remove any residual bandpass structure and then summed over the burst duration, taken to be the time at which the intensity drops to $10\%$ of the peak.

For the CHIME/FRB observations, we follow the procedure of \citet{nimmo2025}. The data are first upchannelised by a factor of 128, yielding a frequency resolution of 3\,kHz, sufficient to resolve the frequency scales measured from the DSA-110 observations. Upchannelisation introduces a periodic scalloping response with a period equal to the native channel width of the CHIME/FRB baseband data: 0.390625\,MHz. We mitigate this artefact by dividing the spectrum at each time sample by a folded off-burst spectrum capturing the scallop shape \citep{nimmo2025}. Just as for the DSA-110 data, the spectra are then normalized by the off-burst standard deviation and then summed over the burst duration to form the burst spectrum.

\subsubsection{Scintillation Bandwidths}
The scintillation bandwidth is measured from the ACF of each burst spectrum, computed following Equation\,2 in \citet{nimmo2025} (where we omit the zero lag noise spike). We fit the ACF as a function of frequency lag, $\Delta\nu$, with a Lorentzian, ${m_{\rm d}^2}/{\left(1+\left(\Delta\nu/\Delta\nu_{\rm d}\right)^2\right)}$, where $m_{\rm d}$ is the modulation index. For the CHIME/FRB ACFs, we include an additional constant offset term to account for residual correlations arising from imperfect scallop removal and other instrumental effects, such as the known 30\,MHz spectral ripple produced by reflections between the mesh and focal line of the telescope \citep{CHIMECollaboration_2022_ApJS}, feed-coupling and imperfect polarization calibration.

\begin{figure}
    \centering
    \includegraphics[width=0.47\textwidth]{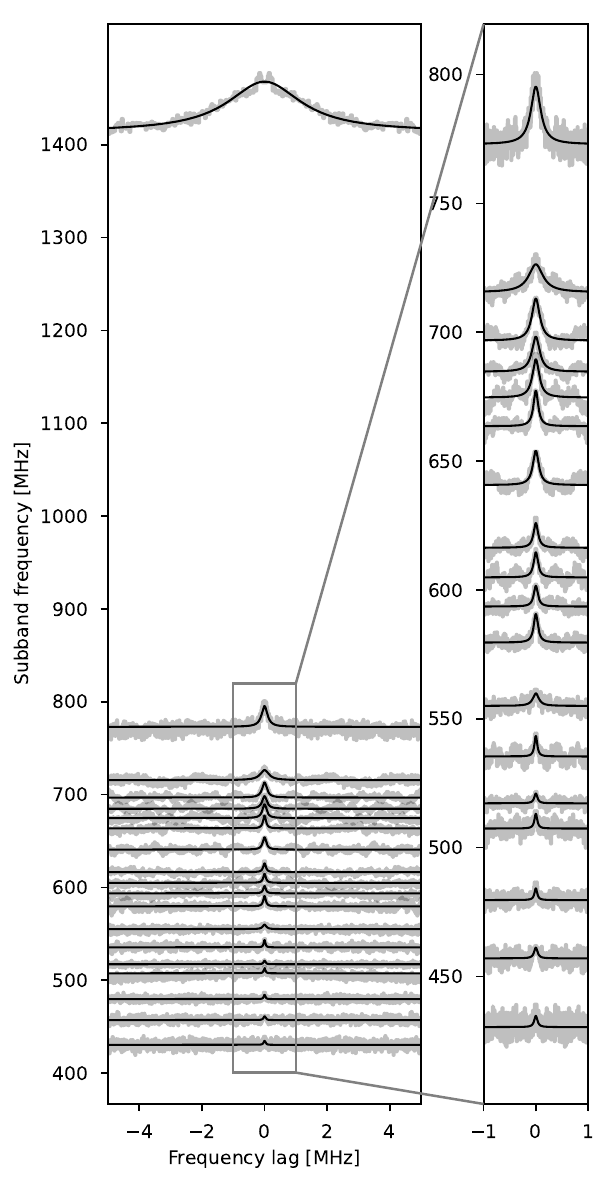}
    \caption{An example of the ACFs measured from CHIME/FRB and DSA-110 spectra for FRB 20240203A/D. Grey curves show the ACFs vs. frequency lag, vertically spaced according to the subbands at which the ACFs were measured and uniformly scaled in amplitude for visibility (the relative height of each ACF therefore reflects differences in the modulation index across subbands). A Lorentzian model for the best-fit scintillation bandwidth is shown in black, for a fitted frequency scaling $\Delta \nu_{\rm d}\propto \nu^{3.82\pm0.23}$ (see text).}
    \label{fig:acf_example}
\end{figure}

The DSA-110 ACF is computed over the full observing band. At the higher observing frequencies of DSA-110, the expected evolution of the scintillation bandwidth across the 200\,MHz bandwidth is modest and dividing the observing band into subbands substantially reduces the number of scintles contributing to each per-subband ACF, increasing the measurement uncertainty. In contrast, the large fractional bandwidth of CHIME/FRB allows the frequency evolution of the scintillation bandwidth to be measured \citep{nimmo2025}. For sufficiently bright bursts, we therefore divide the CHIME/FRB band into multiple subbands and measure $\Delta\nu_{\rm d}$ independently in each. An example of the ACFs measured in multiple frequency subbands is shown for FRB 20240203A/D in Figure~\ref{fig:acf_example}.

The error on any individual scintillation bandwidth measurement is taken to be the quadratic sum of the non-linear least squares fitting uncertainty (i.e., the covariance matrix), the frequency resolution of the spectrum, and the finite scintle error that arises from the limited number of scintles within the chosen frequency bandwidth of the ACF. The finite scintle error is estimated as $\sigma_{\rm fse} \approx \Delta \nu_{\rm d}/\sqrt{N_d}$ where $N_d$ is the number of scintles, given as $N_d = 1+(\eta B/\Delta \nu_{\rm d})$, where $B$ is the observing bandwidth and the scintle filling factor $\eta\approx0.2$ being a typical assumption \citep{cordes_shannon2010,levin2016,brown2025}.

Combining the multi-band CHIME/FRB measurements with the full-band DSA-110 measurement, we fit the expected power-law relation, $\Delta\nu_{\rm d}\propto\nu^\alpha$. Measuring this frequency dependence is an important verification that the observed spectral structure is produced by scintillation rather than self-noise or instrumental systematics, which are not expected to show the same frequency evolution. 

\begin{figure*}
    \centering
    \includegraphics[width=0.95\textwidth]{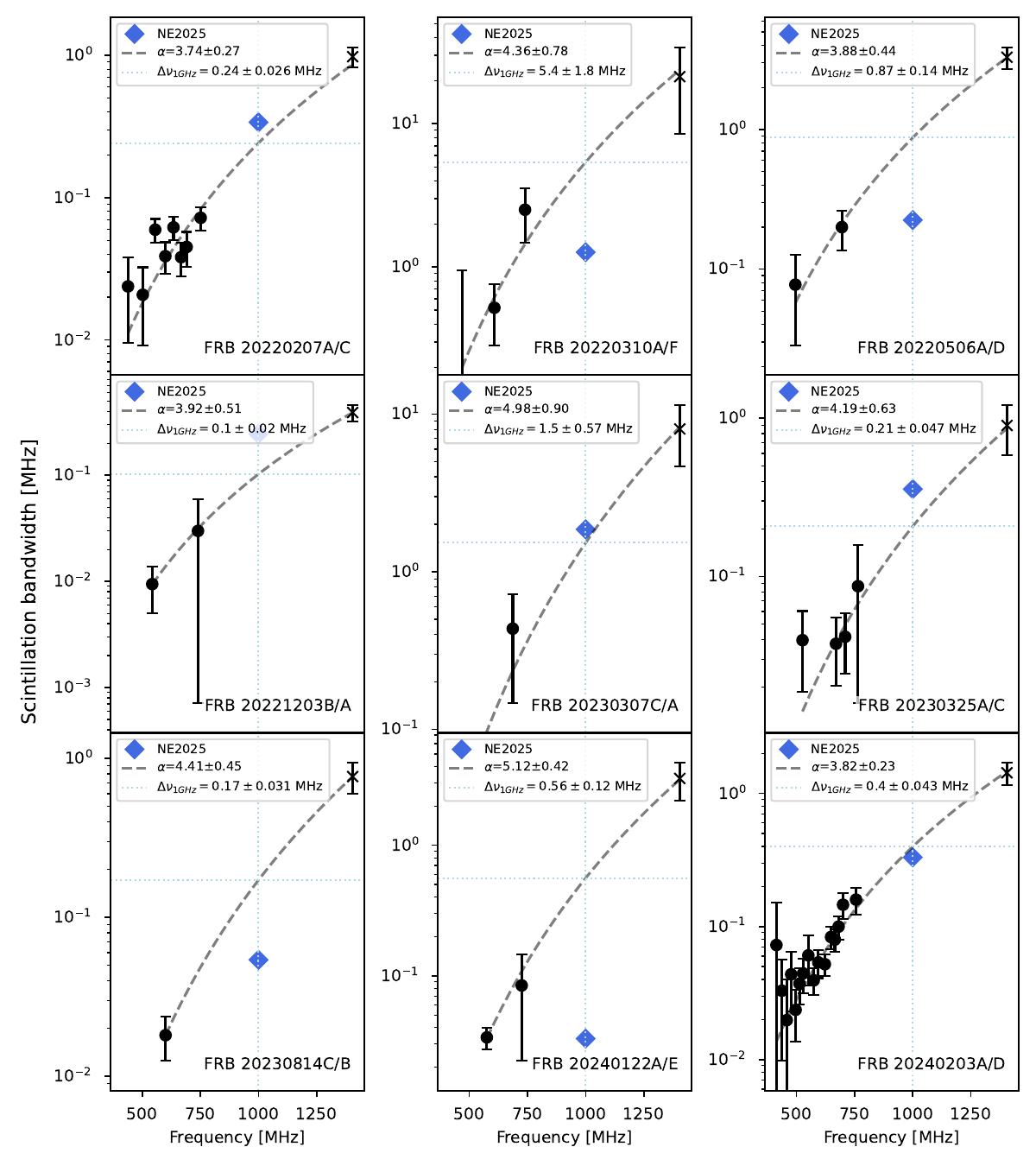}
    \caption{Measured scintillation bandwidths as a function of observing frequency. Black circles show measurements from CHIME/FRB data, where some bursts had sufficient S/N to measure the scintillation across multiple subbands. The black cross denotes the single DSA-110 measurement. Error bars indicate the 1\,$\sigma$ measurement uncertainties added in quadrature with the finite scintle error. The black dashed line shows the best-fit power-law relation, $\Delta\nu_{\rm d}\propto\nu^{\alpha}$.  
    The blue dashed lines show the corresponding scintillation bandwidth at 1\,GHz, and the blue diamonds represent the predicted values for the Milky Way ISM from NE2025 \citep{Ocker_2026_ApJ}. FRB~20221113E/A, 20230913F/G, and 20240229E/A are omitted since no convincing scintillation bandwidth could be measured from the CHIME/FRB and/or DSA-110 observations.}
    \label{fig:scint}
\end{figure*}

\begin{figure}
    \centering
    \includegraphics[width=0.46\textwidth]{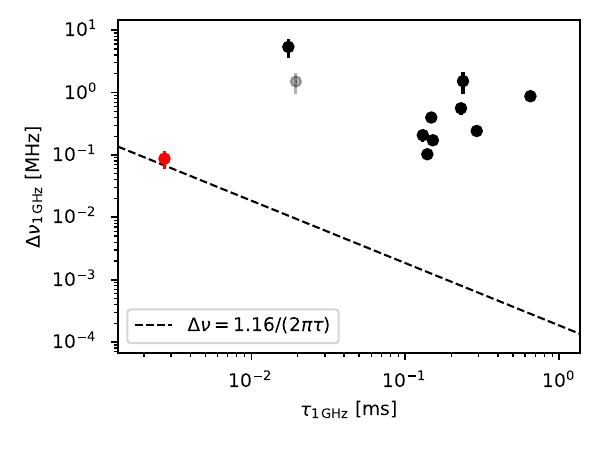}
    \includegraphics[width=0.46\textwidth]{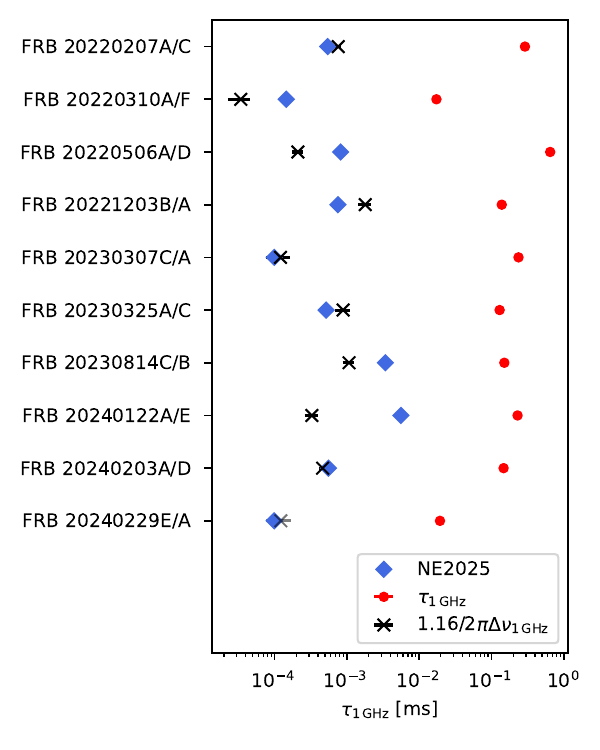}
    \caption{Top panel: Scintillation bandwidth vs. scattering time at 1 GHz. The expected relation between the two for a single scattering screen is shown by the black dashed line. FRB 20230913F/G and 20240229E/A, which lack robust scintillation bandwidth measurements (see text), are highlighted in red and grey, respectively.  
    Bottom panel: Comparison between the scattering times measured directly from pulse broadening (red circles), those inferred from the scintillation bandwidth (black crosses), and the NE2025 Galactic predictions (blue diamonds).
    }
    \label{fig:scattscint}
\end{figure}

We convincingly measure a spectrally evolving scintillation bandwidth in $9$ co-detected FRBs (Figure\,\ref{fig:scint}). In each case, $\Delta \nu_{\rm d}$ extrapolated to 1\,GHz (using the measured scaling index) implies a scattering time inconsistent with the directly measured $\tau$ (Figure\,\ref{fig:scattscint}), using the scaling between scattering and scintillation of a smooth Kolmogorov medium $\tau = 1.16/(2\pi\Delta\nu_{\rm d})$ \citep{cordesrickett98}. The scintillation and pulse broadening therefore originate from distinct screens along the LOS. These scintillation and pulse broadening measurements are reported in Table\,\ref{tab:burst-props}. 

For FRB~20240229E/A, we measure $\Delta\nu_{\rm d}=5.9\pm2.2$\,MHz in the DSA-110 data. Splitting the band in two gives $8.9\pm5.0$ MHz at 1452\,MHz and $6.0\pm2.9$ MHz at 1358\,MHz, broadly consistent with $\Delta \nu_{\rm d}\propto\nu^{4.4}$, but with too few scintles to exclude no evolution. The CHIME/FRB data show no clear frequency scale beyond residual upchannelisation artefacts. The DSA-110 measurement is consistent with the NE2025 prediction for this LOS \citep{Ocker_2026_ApJ}. We show it with reduced opacity in Figure\,\ref{fig:scattscint} because it cannot be validated at a second frequency. 

For FRB~20230913F/G, we measure $\Delta\nu_{\rm d}=0.34\pm0.11$\,MHz (DSA-110), and $0.066\pm0.018$\,MHz (CHIME/FRB), implying $\alpha=1.9\pm0.5$. This is shallower than the expected $\alpha=4.4$, and no evolution is seen across the two CHIME/FRB subbands. Together these suggest that the CHIME/FRB spectral structure may be dominated by self-noise. Scaling the DSA-110 value to 1\,GHz with $\alpha=4$ implies a scattering time of $\sim2.1\,\upmu$s, in agreement with the directly measured value (Table\,\ref{tab:burst-props}; Figure\,\ref{fig:scattscint}). We therefore find no evidence for two distinct screens toward this FRB. Finally, for FRB~20221113E/A, the S/N is too low to measure scintillation in either dataset.

\subsubsection{Modulation Indices}\label{sec:mod-inds}

The Lorentzian fit to the ACF also constrains the modulation index, $m_{\rm d}$, in the same frequency subbands used for the scintillation bandwidths. We have not been able to establish the reliability of these measurements and regard them as preliminary. We report them because $m_{\rm d}$ enters directly into our screen-distance constraints (Section\,\ref{sec:screenlocs}).

In all bursts with a clear scintillation signature, $m_{\rm d}$ evolves with frequency. It increases towards higher frequencies in five FRBs, decreases in three and varies non-monotonically in one. To estimate  $m_{\rm d}$ at 1\,GHz for the screen constraints, we parametrize this evolution as $m_{\rm d}(\nu)=m_{\mathrm{d,1\,GHz}}(\nu/1\,{\rm GHz})^\gamma$, although this form is not physically motivated (see e.g. \citealt{pradeep2025}). Bandpass variations should not produce these trends, since the spectra are normalized by the off-burst statistics. The intrinsic burst spectrum should not produce it either, since each subband is defined to contain equal burst flux. However, we cannot exclude residual instrumental contributions \citep[e.g.][]{nimmo2025}, which we model only as a constant offset in the ACF. 

Within the CHIME/FRB band, $m_{\rm d}$ is anti-correlated with flux density for most bursts, raising the possibility of a S/N-dependent bias. However, $m_{\rm d}$ is unchanged when we artificially degrade the fluence of FRB~20240203A/D to $30$\% of its original value. The same analysis applied to PSR~B1508$+$55 yields the opposite trend (Pradeep et al. in prep). We thus do not interpret the modulation indices further here. In Section\,\ref{sec:screenlocs} we constrain screen distances using both $m_{\rm d}=1$ and, more conservatively, the measured values.

\section{Deconstructing the Sightline}\label{sec:screens}

\subsection{DM Budget}
The DM budget used in this work is based on \cite{pandhi2025} and is also described in Paper II; we briefly summarize it here. We assume the total observed DM is the sum of contributions from the Milky Way interstellar medium (ISM) and halo, the intergalactic medium (IGM), and the host galaxy:
\begin{multline}
    \DM_{\rm obs} = \DM_{\rm MW,ISM} + \DM_{\rm MW,halo} + \DM_{\rm IGM}(z) \\ + \DM_{\rm host}/(1+z)
\end{multline}
where $\DM_{\rm MW,ISM}$ is given by a uniform distribution centered on the NE2025 Galactic electron density model prediction $\pm 20\%$ \citep{Ocker_2026_ApJ}, $\DM_{\rm MW,halo}$ is a log-normal distribution with mean $\mathrm{log}_{10}(30~\mathrm{pc}~\mathrm{cm}^{-3})$ and standard deviation of $0.2$~dex (based, e.g., on \citealt{2019MNRAS.485..648P,cook2023,ravi2025}), and $\DM_{\rm IGM}$ is described by the probability density function of \cite{baptista2024}. The resulting host galaxy DMs in the source's rest-frame are shown in Table~\ref{tab:burst-props}. 

\subsection{Scattering Screen Locations}
\label{sec:screenlocs}

\begin{deluxetable*}{c C C C C C}
    \label{tab:screen-dist}
\tablecaption{Scattering Screen Distances}
\tabletypesize{\footnotesize}
\tablehead{
\rule{0pt}{3ex} &  & \multicolumn{2}{c}{$m_{\rm d}=1$} & \multicolumn{2}{c}{$m_{\mathrm{d,1\,GHz}}$} \\
\cline{3-4} \cline{5-6}
 \colhead{Name} & \colhead{$L_G$ (kpc; NE2025)} & \colhead{$L_XL_G$ (kpc$^2$)} & \colhead{$L_X$ (kpc)} & \colhead{$L_XL_G$ (kpc$^2$)} & \colhead{$L_X$ (kpc)}
}
\startdata
20220207A/C & 2.4 & \lesssim1.1 & \lesssim0.47 & \lesssim2.5 & \lesssim1.0\\
20220310A/F & 1.2 & \lesssim29320 & \lesssim23523 & \lesssim49695  & \lesssim39869 \\
20220506A/D & 2.5 & \lesssim63 & \lesssim25 & \lesssim100&\lesssim40 \\
20230307C/A & 1.1 & \lesssim249 & \lesssim226 & \lesssim674 & \lesssim611 \\
20230814C/B & 2.6 & \lesssim129 & \lesssim50 & \lesssim265 & \lesssim103 \\
20240203A/D & 2.4 & \lesssim10 & \lesssim4.4 & \lesssim14.5 & \lesssim6.1\\
\enddata
\tablecomments{The screen constraints for FRB20220207A/C assume the DSA-110 localization and host galaxy association. The $m_{\rm d}=1$ columns assume unit modulation index; the $m_{\mathrm{d,1\,GHz}}$ columns use each burst's measured modulation index at 1\,GHz (Section~\ref{sec:mod-inds}).}
\end{deluxetable*}

As shown in Section~\ref{sec:scattering-analysis}, the pulse broadening and scintillation measurements of most FRBs in the sample are inconsistent with the uncertainty relation $C_1 = 2\pi\tau\Delta \nu_{\rm d}$, implying the presence of multiple scattering screens along the LOS. Given that all of the scintillation bandwidths are less than a factor of 10 from the NE2025 prediction for each LOS, with the exception of FRB~20240122A/E which is a factor of $\sim17$ lower, we interpret these measurements using a two-screen model in which the scintillation arises from the Milky Way, at some distance $L_G$ from the observer, and the pulse broadening arises from an extragalactic screen, located at some distance $L_X$ from the FRB. Since pulse broadening from an extragalactic (background) screen can quench scintillation by reducing the coherence of emission hitting the foreground scintillation screen, the detection of both of these scattering effects jointly constrains the product of the two screen distances $L_X L_G$ \citep{2015Natur.528..523M,2019ARA&A..57..417C}.

Following \cite{pradeep2025}, the product of the screen distances is
\begin{equation}
    L_XL_G \lesssim \frac{(1+z_s)D_s^2}{8\pi\nu^2}\frac{\Delta \nu_{\rm d}}{m_{\rm d}\tau}
\end{equation}
where $z_s$ and $D_s$ are the redshift and angular diameter distance of the FRB, respectively, and $m_{\rm d}$ is the modulation index of the scintillation. While we measure modulation indices for our sample, determining their robustness will require future work (see Section\,\ref{sec:scattering-analysis}). Therefore, we report screen-distance limits computed both assuming $m_{\rm d}=1$ and using our measured $m_{\rm d}$ at 1\,GHz.  We note that $m_{\rm d} < 1$ is expected if scintillation is significantly quenched by extragalactic pulse broadening, implying that upper limits on $L_XL_G$ inferred using $m_{\rm d}=1$ may be underestimated.

Table~\ref{tab:screen-dist} shows the scattering screen distances, for both assumed modulation indices, for the 6 FRBs that have both scintillation and pulse broadening constraints as well as known redshifts.\footnote{We include scattering screen distance constraints for FRB20220207A/C assuming the DSA-110 localization and host galaxy association.} Upper limits on $L_X$ are calculated using the effective Galactic scattering screen distance predicted by NE2025 \citep{Ocker_2026_ApJ}. Five of the FRBs have upper limits on $L_X$ that are less than the expected sizes of their host galaxy halos (based on their virial radii) for $m_{\rm d}=1$; that number drops to 4 FRBs when adopting the measured $m_{\rm d}$. One of the FRBs has an ambiguously large upper limit: FRB 20220310A/F has $L_X \lesssim 40$ Mpc (for the measured $m_{\rm d}$). An initial search for foreground galaxies along the LOS in existing catalogs (e.g., PanStarrs and DESI; \citealt{chambers2016,desi2026}) does not yield any intervening galaxies near the upper limit on $L_X$ that could plausibly account for the scattering. Given that extensive observational and model-based evidence suggests scattering in the circumgalactic medium is typically negligible \citep{ocker2021halo,masribas2025,ocker2025,jow2026}, the host galaxy remains a plausible source of extragalactic scattering even for this FRB. A detailed investigation of the foreground galaxy distribution for all of these FRBs remains a subject for future work.

\subsection{Scattering in Host Galaxies}\label{sec:host-scattering}

\begin{figure}
    \centering
    \includegraphics[width=0.48\textwidth]{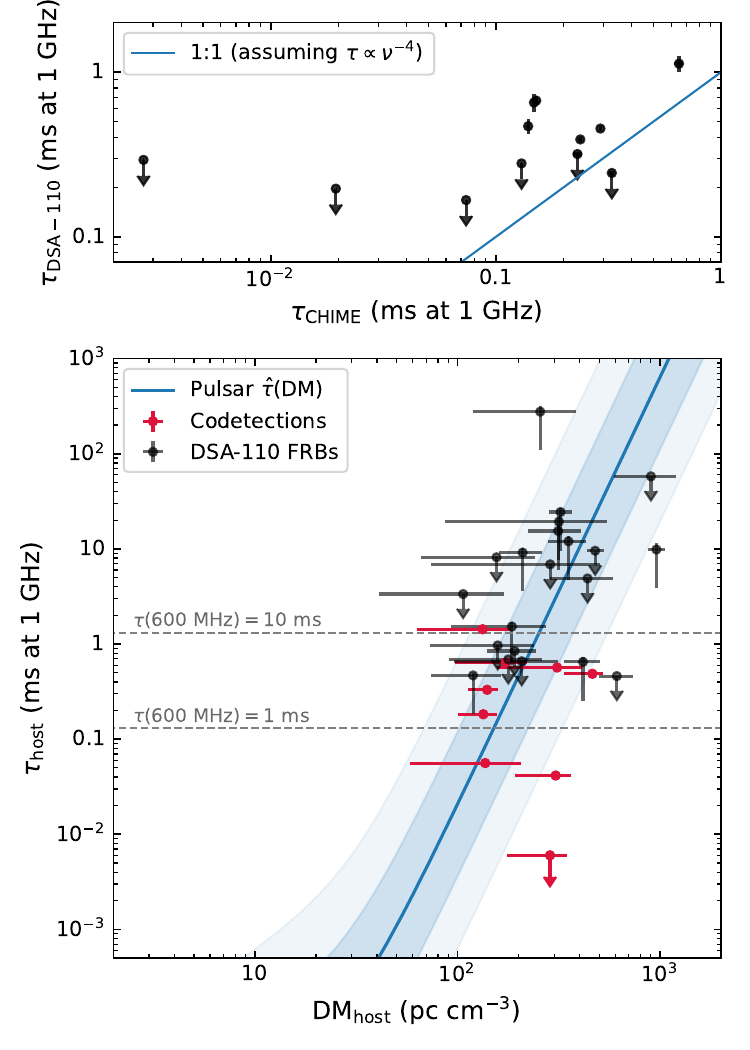}
    \caption{Top panel: Comparison between pulse broadening times $\tau$ fit to DSA-110 data by \cite{verdi2026} and to CHIME/FRB data in this work, both scaled to 1 GHz assuming $\tau\propto\nu^{-4}$. The blue curve shows where $\tau_{\rm DSA-110}=\tau_{\rm CHIME}$. Bottom panel: Pulse broadening time vs. host galaxy DM, both in the host rest-frame, for co-detected FRBs in this work (red) and other DSA-110 FRBs with pulse broadening and redshifts measured by \citealt{verdi2026} (black). Vertical error bars on the DSA-110 FRBs' $\tau$ indicate how much $\tau$ may be overestimated from the DSA-110 data alone, based on the comparison between independent CHIME/FRB and DSA-110 measurements in the upper panel (see text). One co-detected FRB, 20230913F/G, is shown as an upper limit because it has pulse broadening consistent with arising from the Milky Way. The blue curve and shaded regions show the empirical $\hat{\tau}(\DM)$ relation measured from Galactic pulsars and its $\pm1\sigma$ to $\pm2\sigma$ uncertainty, from \cite{cordes2022}. The empirical $\hat{\tau}(\DM)$ has been multiplied by a factor of 3 to account for the difference between spherical and plane wave scattering \citep{ocker2021halo}.}
    \label{fig:tau-DMh}
\end{figure}

The pulse broadening of the co-detected FRBs is likely dominated by their host galaxies, in line with previous studies of independent FRB samples \citep{ocker2022,cassanelli2023,sammons2023,shin2025,glowacki2025}. Whether this scattering arises in the diffuse host ISM or in the FRBs' local environments remains unclear. Figure\,\ref{fig:tau-DMh} compares $\tau$ and host DM, both in the host rest frame, for the co-detected FRBs and the broader DSA-110 sample of FRBs \citep{verdi2026}. It also shows the empirical Galactic pulsar $\tau$--DM relation \citep{cordes2022}, scaled by a factor of three for plane-wave (FRB) rather than spherical-wave (pulsar) sources. An exact correspondence with the Galactic relation is not expected, given its $0.76$ dex scatter and its dependence on Galactic structure and screen geometry \citep{ocker2021halo, ocker2024c}. However, some $\tau$--DM correlation should exist within FRB hosts if both arise in the same medium \citep{cordes2016, ocker2022b}. 

The co-detected FRBs, which have measurable pulse broadening at CHIME/FRB frequencies but not often at 1.4\,GHz, fill out the low-$\tau$, low-DM$_{\rm host}$ region of this parameter space. Their spread is similar to that of the Galactic relation. FRB~20230913F/G has an unusually small $\tau$ for its host DM, which we discuss further in Section~\ref{sec:disc-ind}. For FRBs with $\tau$ measured independently in both CHIME/FRB data (this work) and DSA-110 data \citep{verdi2026}, the DSA-110 values are systematically larger, with a median ratio $\tau_{\rm DSA-110}/\tau_{\rm CHIME} \approx 2.5$ (top panel of Figure\,\ref{fig:tau-DMh}). Both analyses assumed an exponential PBF with $\tau\propto\nu^{-4}$. The difference is likely due to the coarser time resolution and narrower bandwidth of DSA-110, and likely affects other instruments operating at similar frequencies and bandwidths (e.g. ASKAP; \citealt{scott2025}). We therefore inflate the error bars of the DSA-110-only $\tau$ measurements by this factor. Spearman-$\rho$ correlation tests of $\tau$ versus host DM give $p=0.26$ for the DSA-110 FRBs of \citet{verdi2026} (excluding upper limits), $p=0.57$ for the co-detected FRBs, and $p=0.03$ for the combined sample. After a Bonferroni correction for the three tests, the combined value is $p\approx0.09$. Resampling $\tau$ and host DM within their uncertainties gives a 95th-percentile value of $p<0.52$. We thus find evidence for a tentative but not yet significant correlation, consistent with the tentative correlation reported in \citet{verdi2026}. 

\begin{figure}
    \centering
     \includegraphics[width=0.45\textwidth]{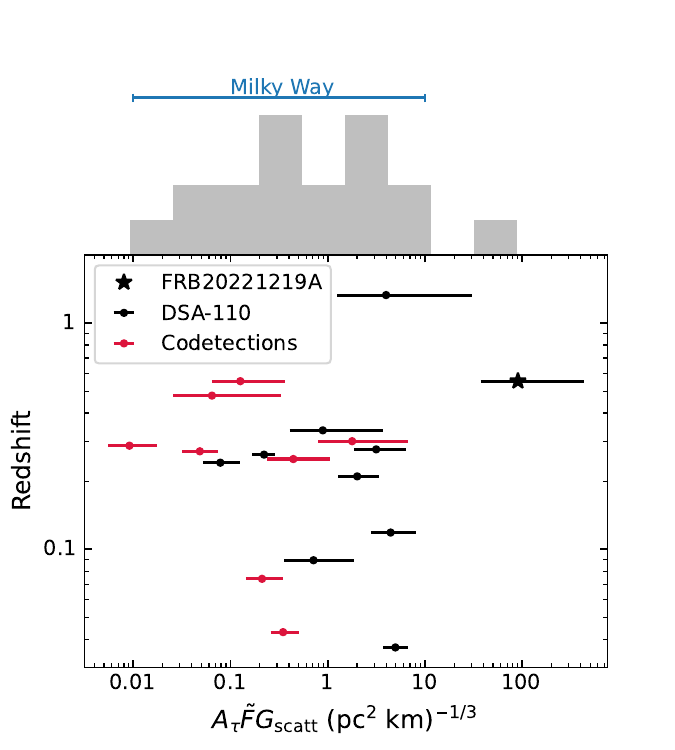}
    \caption{Redshift vs. the composite density fluctuation strength $A_\tau \tilde{F} G_{\rm scatt}$ (see Section~\ref{sec:host-scattering} for details). Pulse broadening measurements for co-detected FRBs based on CHIME/FRB data are shown in red, measurements based on DSA-110 data from \cite{verdi2026} in black, and one FRB outlier is highlighted with a black star. A histogram of $A_\tau\tilde{F}G_{\rm scatt}$ for both samples combined is shown in the top panel.}
    \label{fig:Ftilde-z}
\end{figure}

The reported correlation between $\tau$-RM$_{\rm host}$ in Paper II and DM$_{\rm host}$-RM$_{\rm host}$ \citep{sherman2023, verdi2026} suggests that $\tau$ and DM$_{\rm host}$ likely arise from the same medium despite the lack of clear correlation so far. If this is true, their ratio constrains the density fluctuation strength, $A_\tau\tilde{F}G_{\rm scatt} \approx \tau_{\rm 1\ \rm GHz}(1+z)^3/(48.03\ {\rm ns}\ \DM_{\rm host}^2)$ \citep{cordes91,cordes2016,ocker2022b,cordes2022}. Here, $\tilde{F}$ is a composite parameter describing the density fluctuation statistics, $G_{\rm scatt}$ is a geometric factor that depends on where the scattering medium is located with respect to the source and observer, and $A_\tau$ accounts for differences in assumed PBFs \citep{ocker2022b,cordes2022,ocker2025}. The combined sample spans $0.01 \lesssim A_\tau \tilde{F} G_{\rm scatt} \lesssim 10$ (pc$^2$ km)$^{-1/3}$ (Figure\,\ref{fig:Ftilde-z}), matching Milky Way sightlines of similar DM and the independent FRB sample of \citet{cordes2022}. The one outlier, FRB~20221219A, hints at scattering beyond its host \citep{faber24}. We see no trend with redshift, although one might be expected from increasing star formation rates \citep{ocker2022b}. Future comparisons between the density fluctuation strength and host galaxy properties with larger, uniform FRB samples will provide valuable tests of the origin of FRB scattering and its evolution with redshift. 

\section{Discussion}\label{sec:discussion}

We have used the combined frequency coverage of DSA-110 and CHIME/FRB to constrain key properties of broadband FRBs, including their fluences, energies, flux densities, and scattering. Here we consider the implications of these results.

\subsection{Spectral indices and turnovers}\label{sec:disc-spectra}
Most of the sample have substantially higher spectral energies (by up to a factor of 10) at 400–800\,MHz and only two decrease towards lower frequencies. The stacked spectrum gives mean and median spectral indices of $\approx-0.4$ and $\approx-0.3$, respectively (Figure\,\ref{fig:comb_spec}). These values are shallower than the 
mean of $-1.6$ (median $-1.4$) found by \citet{macquart2019} and the $-{1.3}_{-0.4}^{+0.7}$ measured by \citet{shin2022}, although still consistent within $3\,\sigma$. It is closer to the mean and median of $-0.91 \pm 0.05$ and $-0.98 \pm 0.05$ for the 62 least-biased bursts in the first CHIME/FRB baseband catalog\footnote{Detected bandwidth $>\,300$\,MHz, distance from the center of the synthesized beam $< 0.5^\circ$, detection S/N $> 12$, DM $< 2000$\,pc cm$^{-3}$, $\tau_\mathrm{scat} < 10$\,ms and intrinsic width $< 4$\,ms.} \citep[][Pleunis et al. in prep]{chime_basecat1}. Requiring detection in both bands likely favors flat spectra. Individually, $\sim$50\% of our bursts have positive indices (Table~\ref{tab:spec_props}, Figure~\ref{fig:ind_spec}), compared with $\sim$35\% in \citet{macquart2019}, which may partly reflect bursts sampled on either side of a spectral turnover. Diffractive or refractive scintillation may also contribute to the burst-to-burst scatter.

This spectral diversity resembles Galactic magnetars. Their time-averaged radio spectra are typically flat \citep[e.g.][]{Camilo_2006_Natur, Lazaridis_2008_MNRAS}, but individual pulses range from steeply negative to positive indices \citep[e.g.][]{Pearlman_2018_ApJ, dai_2019}. Rotation-powered pulsars, by contrast, have predominantly steep negative spectra \citep{Jankowski_2018_MNRAS}, including for single pulses \citep{kramer2003,janagal2023}.

Four FRBs show evidence for a turnover between the CHIME/FRB and DSA-110 bands, at $\approx$720--790\,MHz. This is consistent with the more positive spectral indices seen by CHIME/FRB relative to ASKAP \citep{macquart2019}, and resembles the broken power-law spectra of the first multi-telescope detections of FRB~20121102A \citep{Law_2017_ApJ}. These peak frequencies may be biased low since the single high-frequency point available per burst leaves the peak frequency and declining spectral index highly degenerate. Some turnovers may also reflect temporally distinct burst components independently detected by CHIME/FRB and DSA-110 (see Section\,\ref{sec:identify_codetect}). Two of the four show CHIME/FRB-DSA-110 time delays inconsistent with zero (Figure\,\ref{fig:toa-diffs}), consistent with this possibility. Others may reflect scintillation modulating the relative flux in the two bands.

If intrinsic to a single burst, these turnovers are at higher frequencies than is typical for pulsars ($100$--$600$\,MHz; \citealt{Sieber_1973_A&A}). They are instead comparable to GHz-peaked-spectrum pulsars \citep{Kijak_2011_A&A}, and the magnetar XTE~J1810$-$197 \citep{Maan_2022_ApJ}. Such spectra occur in only $\sim3\%$ of pulsars \citep{Jankowski_2018_MNRAS}, compared to $\sim33$\% of our sample. Given our small selection-biased sample of single bursts, however, we draw no conclusion about the prevalence of such peaks among FRBs. Gigahertz peaks in pulsars are commonly attributed to free-free absorption from dense, ionized gas in the local environment, such as pulsar wind nebulae or \HII regions \citep{Kijak_2011_MNRAS}, or in plasma ejected during outbursts \citep{Lewandowski_2015_ApJ}; similar propagation effects in FRB source environments could produce a spectral cutoff \citep{ravi_loeb_2019}. In XTE~J1810$-$197, the spectral peak evolved with time, suggesting that $\nu_{\rm peak}$ may itself trace the source age (\citealt{Maan_2022_ApJ}; see also \citealt{Kijak_2011_A&A}). 

Alternatively, the turnover may be intrinsic to the FRB emission process. Synchrotron maser emission from a decelerating relativistic blast wave predicts a narrowband spectrum peaking at $\sim0.1$--$10$\,GHz, with $F_\nu \propto \nu^{-2}$ above the peak and an induced Compton suppression below it \citep{Metzger_2019_MNRAS, Plotnikov_2019_MNRAS}. Our peak frequencies fall within this range. All three bursts with constrained $\beta_2$ are consistent with $-2$ within $2\sigma$. The steep rising indices of FRBs~20230307C/A and 20230325A/C ($\beta_1\approx2.4$--3.1) are consistent with a low-frequency cutoff. However, emission from a relativistic source larger than its Doppler beaming angle, as expected for shocks beyond the magnetosphere, must have $\Delta\nu_{\rm FWHM}/\nu_{\rm peak} \geq 0.58$ \citep{Kumar_2024_ApJ}, or $\geq 0.18$ \citep{beloborodov_2026}. Our broken power-law fits imply $\Delta\nu_{\rm FWHM}/\nu_{\rm peak} \approx 0.7$--$1.1$, although at least one half-power point lies outside the observed bands. None of our fits violate this limit, so the spectra alone do not discriminate between magnetospheric and far-away emission models.

\subsection{Scattering}\label{sec:disc-scatter}

Most of the codetected FRBs show evidence of both pulse broadening and scintillation, and nine show evidence for two distinct scattering screens along the LOS. Combining the DSA-110 and CHIME/FRB scintillation measurements yields precise constraints on the frequency dependence of the scintillation bandwidth: $\alpha\approx3.7$--$5.1$. Most are consistent within $1\sigma$ with $\alpha=4$ or $4.4$, expected for Gaussian or Kolmogorov-like scattered images. The only exception is FRB~20240122A/E, for which $\alpha$ is $>2\sigma$ above $\alpha=4$. We see no evidence for shallow scattering spectral indices, which are expected in the regime of two-screen scattering when the screens partially or fully resolve each other \citep{pradeep2025}. Comparison with NE2025 \citep{Ocker_2026_ApJ} suggests the scintillation arises in the Milky Way, and hence the pulse broadening arises closer to the source.

Combined with the DSA-110 scattering measurements (18 FRBs total; \citealt{verdi2026}), we find tentative evidence of a correlation between pulse broadening time and host galaxy DM ($p=0.09$; Section\,\ref{sec:host-scattering}). This falls short of a $3\sigma$ detection and may reflect our still-small sample size, so we encourage future studies with larger samples to test its validity. Previous studies disagree on whether such a correlation exists: \citet{verdi2026} report marginal evidence in 57 DSA-110 FRBs (27 with scattering upper limits), while \citet{scott2025} find none in 29 ASKAP FRBs. \citet{masribas2026} argue that, even if an intrinsic relation exists, its expected scatter combined with the large uncertainties in host DM would wash out any detectable correlation. A growing number of highly scattered FRBs also appear to be significant outliers from the empirical $\tau$-DM relation for pulsars \citep{faber24,shin2025,jahnschindler2026}, although this does not serve as conclusive evidence for or against a $\tau$-DM relation in FRB hosts. The implied density fluctuation strengths are consistent with the Milky Way ISM, most closely resembling pulsar sightlines through the disk and towards the inner Galaxy. This may hint that FRBs preferentially reside in dense regions of their hosts such as near spiral arms or star-forming regions \citep{ocker2022b,verdi2026}), although FRB sightlines cross their hosts at arbitrary inclinations.

Finally, $\tau$ measured from DSA-110 data alone is systematically larger than from the CHIME/FRB data (Figure~\ref{fig:tau-DMh}). This suggests that scattering measurements obtained at $\gtrsim1$\,GHz frequencies may be biased high, which may partly explain differences between those reported here and the ASKAP $\tau$ distribution \citep{scott2025}. Combining DSA-110 and CHIME/FRB data probes host galaxy scattering across a much broader range of $\tau$ than either telescope alone. We discuss the role of scattering as a selection bias further below.

\subsection{Selection Effects}
\label{sec:selection}

\begin{figure}
    \centering
    \includegraphics[width=0.45\textwidth]{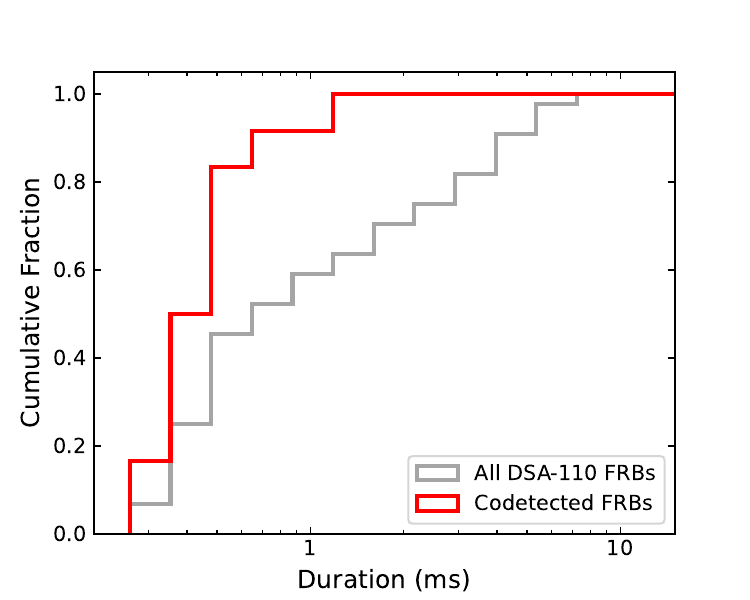}
    \caption{Total duration of co-detected FRBs (red) at 1.4 GHz compared to all DSA-110 FRBs (grey).}
    \label{fig:dsa-widths}
\end{figure}

\begin{figure}
    \centering
    \includegraphics[width=0.47\textwidth]{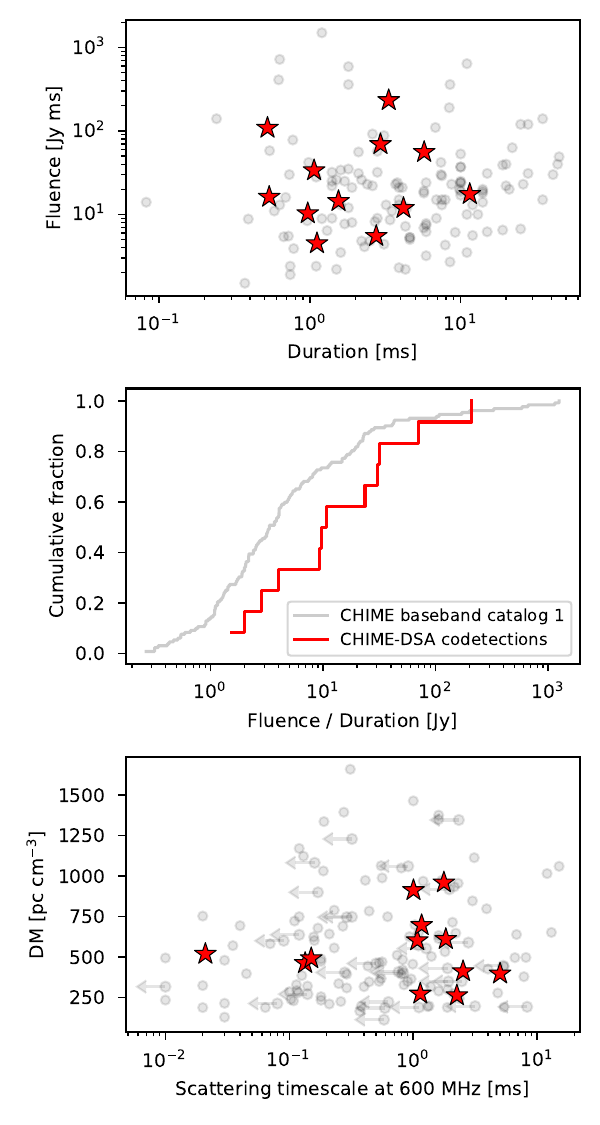}
    \caption{Comparison between the fluence, time duration, average flux density, scattering timescale and DM of co-detected FRBs (red stars and histogram) and FRBs in the first CHIME/FRB baseband catalog (grey points and histogram) \citep{chime_basecat1, sand2025}. Top panel: Fluence vs. total time duration, taken to be the $90\%$ burst width. Middle panel: Cumulative distribution function for the average flux density (fluence/duration). Bottom panel: Scattering timescale measured at 600\,MHz vs. DM. 
    }
    \label{fig:chime_sample_comparison}
\end{figure}

We next consider whether the co-detected FRBs are representative of the broader FRB population. At 1.4\,GHz they are narrower than the overall DSA-110 sample (Figure~\ref{fig:dsa-widths}).  The median co-detected FRB corresponds to the 24th percentile of the DSA-110 total burst duration distribution, and the 5th percentile of the DSA-110 scattering time distribution, though the latter are biased high (Section\,\ref{sec:host-scattering}). At 600\,MHz, however, the co-detections lie at the more scattered end of the CHIME/FRB baseband sample. The median scattering timescale is $1.15^{+1.16}_{-1.01}$\,ms, about twice the catalog median of $0.55^{+1.67}_{-0.43}$\,ms.\footnote{Scattering upper limits are taken at face value; Figure~\ref{fig:chime_sample_comparison} shows most limits lie below the bulk of the co-detections. As an extreme test, setting all limits below the shortest co-detected scattering timescale lowers the catalog median to $0.25$\,ms, strengthening the offset.}. Nine of the twelve co-detections lie above that median. The median co-detection lies at the 34th percentile in duration but the 68th percentile in scattering time, while fluence and average flux density are typical of the catalog (Figure~\ref{fig:chime_sample_comparison}). Scattering thus makes up an unusually large fraction of their 600\,MHz durations, consistent with intrinsically narrow emission at 1.4\,GHz, where scattering is low.

CHIME/FRB becomes severely biased against FRB detection for $\tau\gtrsim10$ ms at 600\,MHz \citep{chimecat1,merryfield2023,mcgregor2026}, corresponding to $\tau \approx 0.25$\,ms at 1.5\,GHz for $\tau\propto\nu^{-4}$, which is the finest resolution of the DSA-110 real-time FRB search. A DSA-110 burst whose width is dominated by scattering is therefore at risk of exceeding CHIME/FRB's scattering limit, so co-detection favors bursts whose 1.4\,GHz duration is set by their intrinsic widths. We thus predict that the co-detection rate will increase if DSA-110 searches at finer time resolution, or if CHIME/FRB's bias against long durations is reduced. The latter is already being realized by CHIME/Slow, a commensal search pipeline sensitive to transients with widths of 16\,ms to 5\,s \citep{Mate_2026_ApJ}.

To test whether other DSA-110 FRBs detected during this period could plausibly have been detected by CHIME/FRB, we compute the minimum detectable CHIME/FRB fluence, $F_{\rm min}$, at the position and arrival time of each of the 35 non-co-detected FRBs in the DSA-110 archive\footnote{\url{https://code.deepsynoptic.org/dsa110-archive/}}. 
The on-axis fluence threshold follows from the radiometer equation for S/N$=8.5$, an SEFD of 45\,Jy \citep{chimefrb_overview}, two polarizations and the full CHIME/FRB bandwidth. We scaled the fluence by the daily system sensitivity measured from Galactic pulsars \citep{Josephy2019}, and by the band-averaged response of the most sensitive operating formed beam at the burst position and time \citep[including sidelobes;][]{andersen2023}. A burst at $F_{\rm min}$ thus has only a $\approx50\%$ detection probability. Since the 600\,MHz width is unknown, we assumed the DSA-110 widths are entirely due to scattering and scaled them as $\nu^{-4}$. This is the maximum scattering the DSA-110 data permit, and it maximises $F_{\rm min}$. We extrapolate each DSA-110 fluence to 600\,MHz with the median stacked spectral index determined from the co-detections, $\beta=-0.3$, which we also used to weight the beam response across the CHIME/FRB band.

Of the 35 non-co-detected DSA-110 FRBs, none are predicted to exceed $F_{\rm min}$ by a significant factor after considering uncertainties in the DSA-110 fluences, the $F_{\rm min}$ and the spectral index\footnote{Six are predicted to exceed $F_{\rm min}$, but only by factors of 1--1.5.}. Under the assumption of scattering-dominated DSA-110 widths, therefore, the CHIME/FRB non-detections do not require a population of FRBs band-limited to frequencies above the CHIME band. The scattering-dominated case is consistent with the selection effect described above, in which co-detections favor bursts whose DSA-110 widths are set by their intrinsic width. In the limit of no scattering, 27 DSA-110 FRBs would exceed $F_{\rm min}$ (15 by more than a factor of $\sim2$), and their non-detection would require positive spectral indices or band-limited emission.  These non-detections could then be explained by the anti-correlation between intrinsic width and spectral occupancy suggested by \citet{pleunis2021}, in which repeating FRBs tend to have longer-duration bursts with more narrowband emission. 

It thus remains unclear whether CHIME/FRB and DSA-110 probe the same underlying FRB population, which depends on how common highly scattered FRBs are. If CHIME/FRB is representative, DSA-110 samples the high-scattering tail; if DSA-110 is representative, CHIME/FRB samples the low scattering tail. The interplay between time resolution and frequency dependent scattering will be important for wideband surveys such as CHORD and DSA, 
which will necessarily employ subbanded searches to minimize RFI contamination and computational cost. If the rate of intrinsically narrow bursts is high, then there should be minimal loss in sensitivity for a search operating at coarser time resolution; conversely, if the burst rate is a shallow or constant function of width, then meaningful losses in sensitivity will be incurred for searches at coarser time resolutions. That only three (or fewer) of our twelve co-detections are minimally scattered at 400-800\,MHz may tentatively favor the former scenario, and is consistent with the narrower burst widths found for non-repeating FRBs (also wider bandwidths) by \citet{pleunis2021} -- though we caution that our sample is too small to draw a firm conclusion. A conservative strategy is to increase the time resolution at higher observing frequencies.

\subsection{Discussion of Individual Sources}\label{sec:disc-ind}

Two co-detected FRBs are notable individually:

\textbf{FRB 20230814C/B} is the only repeating FRB in the co-detected sample. It has been detected once more by the DSA-110, 47 days before the codetection \citep{johndoe_atel} but not by any other telescope thus far\footnote{The positions and DMs are consistent to $<1\sigma$.}. It is also currently the only repeater detected by both CHIME/FRB and the DSA-110 (a possible exception is FRB~20220521B; Section~\ref{sec:identify_codetect}). Its emission spans the entire frequency bandwidths of both DSA-110 and CHIME/FRB, with some evidence for two burst components in the DSA-110 frequency-averaged profile. It shows no downward-drifting sub-bursts, a common hallmark of repeating FRBs \citep[e.g.][]{hessels2019,pleunis2021,limaye2026}, although drifting outside the DSA-110 band cannot be excluded. Its scattering timescale and screen-distance constraint are unremarkable. Its $\mathrm{RM}_{\rm host}=39^{+45}_{-47}$\,rad\,m$^{-2}$ (Paper II), is far below the extreme values of repeaters with compact persistent radio counterparts \citep[e.g.][]{Michilli_2018_Natur, annathomas2022}, but typical of most repeaters \citep[e.g.,][]{Pandhi_2024_ApJ, Ng_2025_ApJ}. 20230814C/B does, however, show depolarization with decreasing frequency (Paper II), a trend also reported for other repeating FRBs \citep[e.g.][]{Fengscience}. Its low apparent burst rate and non-repeater like emission properties could fit a picture in which repeating and non-repeating FRBs share a source population with a range of emission activity, although one source is not conclusive.

\textbf{FRB 20230913F/G} has the smallest inferred scattering time in the sample, $\tau = 2.722\pm0.006~\mu$s at 1\,GHz. This value was fit to a leading sub-component of the burst and is consistent with the predicted Milky Way contribution (Section~\ref{sec:scattering-analysis}), so we treat it as an upper limit on host scattering. Yet its redshift ($z=0.3024$) and DM ($\DM=518$ pc cm$^{-3}$) imply $\DM_{\rm host}=286^{+59}_{-111}$ pc cm$^{-3}$, and it has $\rm RM_{\rm host} = -163^{+25}_{-26}$ rad m$^{-2}$ (Paper II). This combination resembles FRB~20121102A, which shows no evidence of scattering from its host galaxy despite having $\DM_{\rm host}\approx200$\,pc\,cm$^{-3}$ and a fluctuating magneto-ionic environment \citep{2018Natur.553..182M,2017ApJ...834L...7T,ocker2021halo}. A highly inhomogeneous circumsource medium could produce strongly variable scattering, so that only the least scattered bursts are detectable \citep{ocker2023}. In the absence of repeat bursts, it is difficult to test this hypothesis for FRB 20230913F/G.\\

\section{Conclusions}
\label{sec:conclusion}

We have presented a sample of $15$ FRBs co-detected by DSA-110 and CHIME/FRB between 2022 and 2024, of which $12$ have baseband data from both instruments and form the basis of our analysis. Our study firmly establishes that non-repeating FRBs can emit across more than 1 GHz of spectral bandwidth.  Previous wideband FRB studies have primarily targeted observations of repeating sources \citep[e.g.][]{Majid_2020_ApJL,kumar2023}, which are well known to typically have narrower spectral bandwidths than non-repeaters \citep{pleunis2021}, although wideband repeat bursts are occasionally seen \citep{Law_2017_ApJ}. All but one of our bursts are as-yet non-repeating FRBs. 

The key results are as follows:
\begin{itemize}
    \item \textbf{Shallower spectra}: The sample-median spectral index is $-0.3\pm0.4$ and the mean is $-0.4\pm0.4$, which are shallower than previous reports \citep{macquart2019,shin2022}, likely due to selection effects.
    \item \textbf{Spectral turnovers}: Some bursts show evidence for a turnover within or between the CHIME/FRB and DSA-110 bands, comparable to the gigahertz-peaked spectra seen in some pulsars and magnetars \citep[e.g.][]{Kijak_2011_A&A}. This may be caused by local absorption, but we cannot rule out scintillation or distinct burst components.
    \item \textbf{Typical energies}: The band-integrated (400--1500\,MHz) energies are $\approx 0.6 -29\times10^{40}$\,erg -- below the lower limit on the FRB energy ceiling reported by \citet{Shah_2026_arXiv}, and smaller than the most energetic FRBs on record ($\gtrsim10^{42}$\,erg assuming 1\,GHz bandwidth; \citealt{Ravi_2019_Natur,Ryder_2023_Sci}).
    \item \textbf{Two-screen scattering}: Nine of twelve bursts show evidence for two scattering screens, via joint scintillation and pulse broadening measurements, with scattering predominantly originating within the host galaxy.
\end{itemize}

We find an interplay between scattering, intrinsic burst width and each survey's time resolution opens only narrow window for co-detected FRBs. The co-detected FRBs are among the narrowest detected by DSA-110, but skewed to larger scattering times at CHIME/FRB. This suggests that the wider DSA-110 bursts are sufficiently scattered to exceed the $\sim10$\,ms scattering selection bias of CHIME/FRB at 600\,MHz. That only three (or fewer) of our twelve co-detections appear to be minimally scattered at 400--800\,MHz tentatively favors a population in which intrinsically narrow, broadband bursts are common. Alternatively, the width--bandwidth dichotomy between repeating and non-repeating FRBs \citep{pleunis2021} may contribute, with intrinsically narrow bursts being broadband and intrinsically wide bursts band-limited. Either way, the two surveys select substantially different parts of the FRB population. Future subbanded wideband surveys such as CHORD and DSA should prioritize increasing time resolution at higher observing frequencies. We predict that ultra-wideband FRBs are likely to be commonplace discoveries for the next generation of FRB surveys. This work offers a pathfinder for future studies, which will ultimately determine whether wideband FRBs are representative of the underlying source population.

\section*{Acknowledgements}

We acknowledge that CHIME is located on the traditional, ancestral, and unceded territory of the Syilx/Okanagan people. We are grateful to the staff of the Dominion Radio Astrophysical Observatory, which is operated by the National Research Council of Canada. CHIME is funded by a grant from the Canada Foundation for Innovation (CFI) 2012 Leading Edge Fund (Project 31170) and by contributions from the provinces of British Columbia, Québec and Ontario. The CHIME/FRB Project is funded by a grant from the CFI 2015 Innovation Fund (Project 33213) and by contributions from the provinces of British Columbia and Québec, and by the Dunlap Institute for Astronomy and Astrophysics at the University of Toronto. Additional support was provided by the Canadian Institute for Advanced Research (CIFAR), McGill University and the Trottier Space Institute at McGill thanks to the Trottier Family Foundation, and the University of British Columbia. The baseband recording system for CHIME/FRB is funded in part by a CFI John R. Evans Leaders Fund award to IHS. This research was enabled in part by software provided by the Digital Research Alliance of Canada (\href{https://alliancecan.ca/}{alliance.can.ca}). 

The authors thank staff members of the Owens Valley Radio Observatory and the Caltech radio group, whose tireless efforts were instrumental to the success of the DSA-110. The OVRO is located on the ancestral homelands of the Big Pine Paiute Tribe of the Owens Valley; we recognize and acknowledge the historical and cultural significance of these lands to members of the Tribe. The DSA-110 is supported by the National Science Foundation Mid-Scale Innovations Program in Astronomical Sciences (MSIP) under grant AST-1836018.

The AstroFlash research group at McGill University, University of Amsterdam, ASTRON, and JIVE is supported by: a Canada Excellence Research Chair in Transient Astrophysics (CERC-2022-00009); an Advanced Grant from the European Research Council (ERC) under the European Union's Horizon 2020 research and innovation programme (`EuroFlash'; Grant agreement No. 101098079); and an NWO-Vici grant (`AstroFlash'; VI.C.192.045).
FRB research at UBC is supported by an NSERC Discovery Grant and by the Canadian Institute for Advanced Research. 

S.~K.~O. acknowledges support from the Brinson Foundation and is a member of the NANOGrav Physics Frontiers Center (NSF award PHY-2020265).
K.~N.~acknowledges support by NASA through the NASA Hubble Fellowship grant \# HST-HF2-51582.001-A awarded by the Space Telescope Science Institute, which is operated by the Association of Universities for Research in Astronomy, Incorporated, under NASA contract NAS5-26555.
A.~P.~is a Trottier Space Institute postdoctoral fellow and a Fonds de recherche du Québec - Nature et technologies (FRQNT) postdoctoral fellow.
S.P.E.T is a Fonds de Recherche du Quebec - Nature et Technologies (FRQNT) doctoral fellow (2007570).
A.~P.~C.~is a Canadian SKA Scientist and is funded by the Government of Canada / est financé par le gouvernement du Canada.
V.~M.~K. holds the Lorne Trottier Chair in Astrophysics \& Cosmology, a Distinguished James McGill Professorship, and receives support from an NSERC Discovery grant (RGPIN 228738-13).
M.~N.~is a Fonds de Recherche du Québec - Nature et Technologies (FRQNT) postdoctoral fellow.
A.~B.~P.~acknowledges support from NASA through the NASA Hubble Fellowship grant \mbox{HST-HF2-51584.001-A} awarded by the Space Telescope Science Institute, which is operated by the Association of Universities for Research in Astronomy, Inc., for NASA, under contract \mbox{NAS5-26555}. A.~B.~P.~also acknowledges prior support from a Banting Fellowship, a McGill Space Institute~(MSI) Fellowship, and a Fonds de Recherche du \mbox{Qu\'ebec -- Nature} et Technologies~(FRQNT) Postdoctoral Fellowship.
M.~W.~S.~acknowledges support from the French government under the France 2030 investment plan, as part of the Initiative d’Excellence d’Aix-Marseille Université - AMIDEX (AMX-23-CEI- 088).
P.~S.~acknowledges the support of an NSERC Discovery Grant (RGPIN-2024-06266).

\textit{Data Availability}: Data will be provided prior to publication in a public archive.

\appendix

\section{Dynamic Spectra for FRBs 20220319D, 20220726B/A, and 20230712H/A}\label{app:extra-codetections}

\begin{figure}[h]
    \centering
    \includegraphics[width=0.7\textwidth]{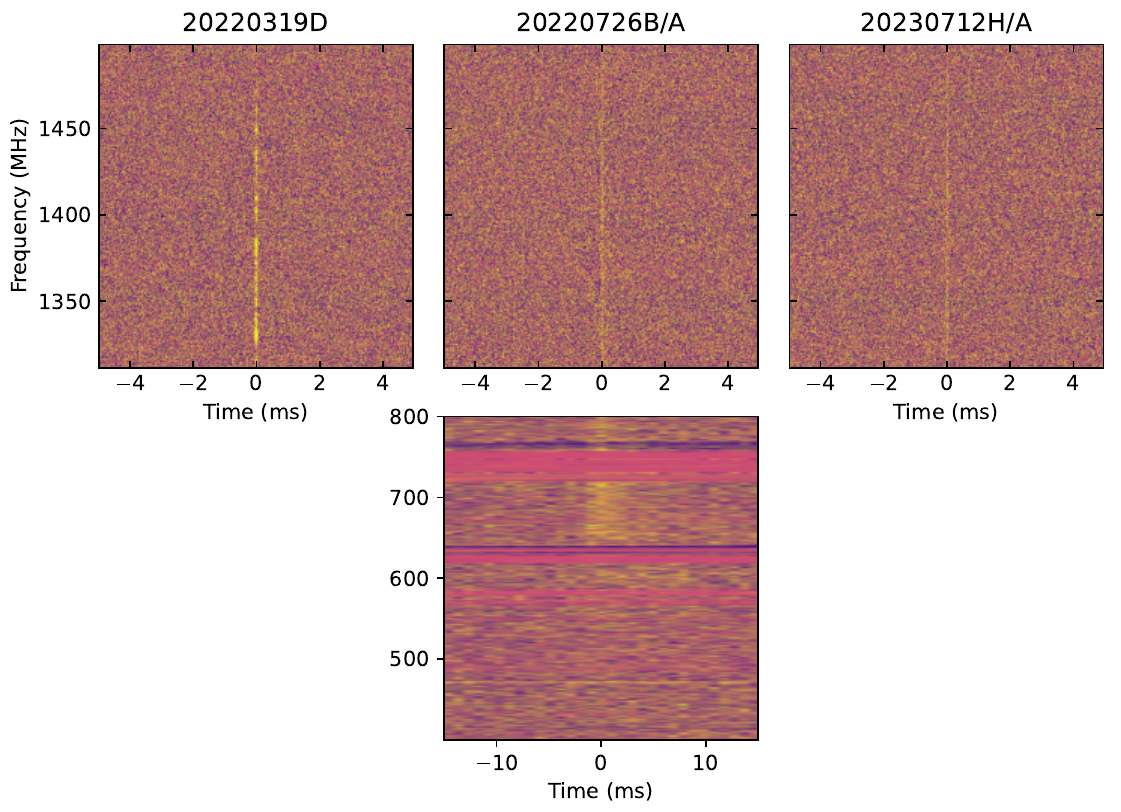}
    \caption{Dynamic spectra for three co-detected FRBs that were not included in the full analysis because they lack CHIME/FRB voltage data. Top panels show the DSA-110 spectra and the bottom panel shows the CHIME/FRB spectrum for FRB 20220726B/A.}
    \label{fig:appendix-ds}
\end{figure}

Figure~\ref{fig:appendix-ds} shows dynamic spectra for the three co-detected FRBs, 20220319D, 20220726B/A, and 20230712H/A, that were not included in our full analysis due to a lack of voltage data from CHIME/FRB. FRB 20220726 was too low S/N to trigger a full voltage dump, and CHIME/FRB experienced some unknown technical error during the time of FRB~20230712, possibly a period of high real-time processing latency, resulting in missing baseband and intensity data. FRB 20220319D was classified as Galactic by the CHIME/FRB real-time search so no voltage dump was triggered.

\section{Localization Comparison for FRB 20220207A/C}\label{app:zach}

\begin{figure}[h]
    \centering
    \includegraphics[width=0.5\linewidth]{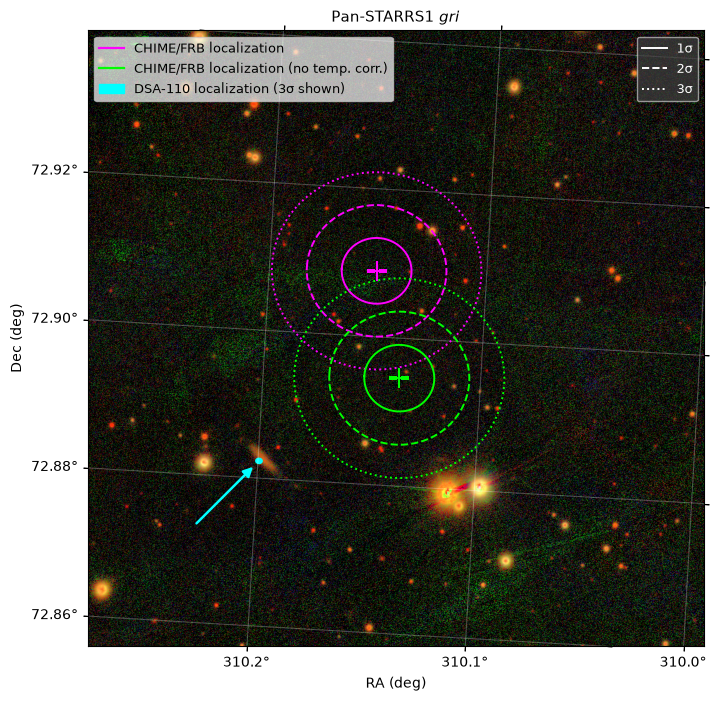}
    \caption{Pan-STARRS1 \citep{chambers2016} composite image showing the offset between the DSA-110 and CHIME/FRB/baseband localizations for FRB 20220207A/C. The size of the DSA-110 localization shown corresponds to the $3\sigma$ uncertainty region. The DSA-110 localization was used to report an association with galaxy PSO J310.1977$+$72.8826, indicated by the blue arrow, in \cite{law2024}. Two CHIME/FRB localizations are shown, one prior to applying a temperature correction (green) and one after applying a temperature correction (magenta); see Appendix~\ref{app:zach} for details.}
    \label{fig:zach-host}
\end{figure}

Figure~\ref{fig:zach-host} shows the localizations for FRB 20220207A/C from DSA-110 and CHIME/FRB/baseband data, superposed on an optical image that was used to identify a host galaxy association in \cite{law2024}. This FRB met all of the initial selection criteria used to identify co-detected bursts (Section~\ref{sec:sample}), including a consistent DM and a barycentric ToA difference consistent with zero, and its initial localization based on CHIME/FRB intensity data was formally consistent with the DSA-110 localization. However, reprocessing of the CHIME/FRB/baseband data, which yields a more precise localization region, results in a position that is formally inconsistent to $>3\sigma$ with the DSA-110. This appendix lays out the various cross-checks that were performed to assess the localizations obtained with each telescope.

\subsection{DSA-110 Cross-Checks}

The latest description of the DSA-110 FRB localization procedure, as applied to FRB 20220207A/C, is given in \citet{sharma2024} (Methods). A summary of this procedure is provided here. The DSA-110 continuously stores full cross-correlation data, phased to the local meridian at the observing declination, and stores triggered voltage dumps for FRBs. The voltage dumps are correlated offline. All cross-correlations are manipulated and analyzed using the  Common Astronomy Software Applications (CASA) package, including the final imaging steps used for localizations. For each FRB, cross-correlation data on a bandpass calibrator observed within 24\,hr (3C309.1 in this case) are phased to the location of the bandpass calibrator for 5\,min around transit, and bandpass calibrations are obtained assuming a point-source model. These calibrations are applied to similar datasets on at least eight VLBI calibrators observed within $\pm2$\,hr of the FRB. The VLBI calibrators are imaged, and position offsets (typically under 2\arcsec) are interpolated to the FRB occurrence time. The FRB cross-correlations are also imaged after bandpass calibration, and the measured source position is corrected using the interpolated offsets. We check the performance of the calibration and position corrections using images of a 10\,min dataset obtained around the FRB time, and the full FRB voltage dump, by comparing measured source positions with cataloged positions from the NRAO VLA Sky Survey (NVSS). We have verified the performance of this localization procedure using voltage dumps on calibrator sources and pulsars. 

Besides verifying the correct performance of the pipeline for FRB 20220207A/C, we performed the following additional checks.
\begin{itemize}
    \item We checked the amplitude gains derived on the bandpass calibrator, and verified that all antennas had similar gains. We verified that there were no unusual offsets in the positions of the VLBI calibrators and NVSS sources (including those detected in the voltage dataset), and a reasonable noise level in the image of the 10\,min dataset around the FRB time. These checks confirm the quality and applicability of the bandpass calibrations given observing conditions at the time. 
    \item We imaged the FRB and the full voltage datasets using the bottom and top halves of the band, and verified that there were no position changes in detected sources. This guards against various types of software errors in phasing. 
    \item We imaged the FRB and the full voltage datasets with each DSA-110 outrigger antenna successively dropped, but saw no more than statistical variations in the derived source positions. This removes the possibility of bad antennas or signal chains affecting the localization. 
    \item We phased the FRB data to the CHIME/FRB position of FRB 20220207A/C, and re-imaged, but found no shift in the DSA-110 position. This negates the possibility of aliasing in the imaging process. We also verified that the CHIME/FRB position is within the DSA-110 primary beam. 
\end{itemize}

We cannot think of any further means whereby the DSA-110 position for FRB 20220207A/C could be incorrect by significantly more than the quoted  errors. 

\subsection{CHIME/FRB Cross-Checks}

When baseband data are saved for a burst, CHIME/FRB can obtain localizations with up to $\sim$sub-arcminute accuracy and precision \citep{michilli2022}.
Well-localized radio sources (including both pulsars and VLBI-localized FRB repeaters) often have baseband data saved as well;
some bursts from these sources are then \textit{ab initio} localized in order to empirically calibrate systematics and uncertainties on the baseband localizations \citep{Michilli_2021_ApJ}.
A more updated version of this systematics calibration has been carried out for bursts that occurred after the time range of the initial baseband pipeline localization calibration.
The updated calibration sample includes 400 bursts from $>$30 unique pulsars and 6 unique FRBs spanning a broad range of dates (2020-01-01 through 2025-09-15), S/N, declinations, and localization offset from telescope meridian.
To account for any time-varying systematics, calibrations were obtained and applied independently in $\sim2$ year intervals.

The result is that for the bursts within that date range, we expect the overall distribution of CHIME/FRB's baseband localizations to be statistically consistent with their ``true'' positions.
This means it is possible (however implausible) for any individual burst to have a localization that is highly statistically inconsistent with its ``true'' localization as a result of short-term systematics that may not have been captured by the calibration sample.
One such example is the quality of the gain calibrations.
In the case of FRB 20191221A, an unusual calibration problem led to a many-degree pointing offset and subsequently the misclassification of a pulsar as an FRB \citep{2026arXiv260607087C}.
While manual inspection of the gains used for FRB 20220207A/C showed they appeared nominal, we re-localized the burst using gains from two different nearby days. 
However, the localizations using both different gains (that also appeared nominal) did not differ significantly from the initial localization.

Another possibility is that the thermal expansion of the telescope was not properly corrected for.
Temperature fluctuations can cause the metal structure of the CHIME telescope to thermally expand or contract, resulting in localization errors on the order of a few arcminutes in the N--S direction.
Temperatures are collected every 5 minutes at the weather station at the Dominion Radio Astrophysical Observatory, the site where CHIME is located.
For every burst localized through the CHIME/FRB baseband localization pipeline, this temperatures database is queried in order to measure a thermal expansion coefficient around the time of the burst detection ($\sim$6 hours before to 6 hours after), which is then used to post facto correct the single-pulse localization.
It is possible that around the time FRB 20220207A/C was detected, the temperatures reported at the DRAO weather station differed significantly from the temperatures at the telescope, although the temperature gradient within the same observatory site would have to be extreme.
It is also possible that the thermal expansion coefficient did not well-describe the temperatures experienced around the time-of-arrival of FRB 20220207A/C;
the temperature corrections increased the declination of the localization by 0.86\,arcmin (i.e., the CHIME/FRB declination was more consistent with the DSA-110 localization before temperature corrections were applied).
Nonetheless, based on our empirical calibrations, we do not generally expect a highly unreliable temperatures-correction procedure to be a common occurrence for CHIME/FRB baseband localizations.

\bibliography{master_bib}

\end{document}